\documentclass[10pt]{article}
\usepackage{amssymb}
\usepackage{amsmath}
\usepackage{amstext}
\begin{document}
\renewcommand{\today}{}
\newcommand{\inv}{^{-1}} \newcommand{\wick}[1]{{:\!#1\!:}}
\newcommand{\NN}{{\mathbb{N}}} \newcommand{\MM}{{\mathbb{M}}} 
\newcommand{\RR}{{\mathbb{R}}} \newcommand{\CC}{{\mathbb{C}}} 
\renewcommand{\SS}{{\mathbb{S}}} \newcommand{\ZZ}{{\mathbb{Z}}} 
\newcommand{\HH}{{\mathcal H}} \newcommand{\LL}{{\mathcal L}}
\newcommand{\bfa}{\hbox{\boldmath$\vec\alpha$}}
\renewcommand{\theequation}{\thesection.\arabic{equation}}
\def\bea{\begin{eqnarray}} \def\eea{\end{eqnarray}}
\def\bra{\langle 0\vert\;} \def\ket{\;\vert 0\rangle}
\def\un{{\rm 1\mkern-4mu I}}
\title{Partial wave expansion and Wightman positivity \\ in conformal
  field theory} 
\author{Nikolay M. Nikolov$^{1,2^*}$, Karl-Henning Rehren$^{2^*}$,
  Ivan T. Todorov$^{1,2^*}$} 
\date{
\bigskip
\begin{itemize}
\item[$^{1}$] {\small Institute for Nuclear Research and Nuclear Energy} \\
{\small Tsarigradsko Chaussee 72, BG-1784 Sofia, Bulgaria}
\item[$^{2}$] {\small Institut f\"ur Theoretische Physik, Universit\"at G\"ottingen,} \\ {\small Friedrich-Hund-Platz 1, D-37077 G\"ottingen, Germany}
\end{itemize}
}

\maketitle
\begin{center}\today\end{center}
\vglue 1cm

\begin{abstract}
A new method for computing exact conformal partial wave expansions is
developed and applied to approach the problem of Hilbert space
(Wightman) positivity in a non-perturbative four-dimensional quantum
field theory model. The model is based on the assumption of global
conformal invariance on compactified Minkowski space (GCI). Bilocal
fields arising in the harmonic decomposition of the operator product
expansion (OPE) prove to be a powerful instrument in exploring
the field content. In particular, in the theory of a field $\LL$ of
dimension 4 which has the properties of a (gauge invariant)
Lagrangian, the scalar field contribution to the 6-point function of
the twist 2 bilocal field is analyzed with the aim to separate the
free field part from the nontrivial part.
\end{abstract}

\vskip10mm

PACS 2003: 11.10.--z, 03.70.+k


\vglue 35mm

\begin{itemize}
\item[$^{*}$] {\small e-mail addresses: \par
mitov@inrne.bas.bg, nikolov@theorie.physik.uni-goe.de \par
rehren@theorie.physik.uni-goe.de  \par
todorov@inrne.bas.bg, itodorov@theorie.physik.uni-goe.de}
\end{itemize}
\newpage
\tableofcontents

\section{Introduction}\label{sec1}

Conformal partial wave expansions \cite{DPPT76,DMPPT77} (an
outgrow of conformal {\it operator product expansions} (OPE)
\cite{FGGP72}) provide a powerful method in conformal quantum field
theory and continue to attract attention (see
\cite{RS88,LR93,DO01,HPR02,NST03,DO04,NST04,NT04}). We derive a new method
for the complete partial wave expansion in closed form, and apply it to
the positive definite 4-point functions of a pair of scalar fields, in
order to approach the non-linear problem of Hilbert space positivity
which is the central remaining problem when all linear properties of
the correlation functions are satisfied. 

Specifically, we study a model generated by a conformal field $\LL$ of
dimension 4 (in four space-time dimensions) which has the properties
of a Lagrangian density \cite{NST03}. Such a model appears to offer the
best chance for constructing a nontrivial non-perturbative local
quantum field theory.    

\smallskip

To set the stage, we begin by a brief survey of relevant results of
earlier work (\cite{NT01,NST02,NST03}; for a recent review -- see
\cite{NT04}). 

\smallskip

It has been understood more than 30 years ago that the exponentiation
of infinitesimal to finite conformal transformations (global conformal
symmetry) in general requires a non-local decomposition of covariant
fields with respect to the center of the conformal group \cite{SV73,SSV75}, 
and that this structure can be conveniently taken into account by the
passage to a covering space \cite{LM75}. 

We here pursue the model assumption called {\em global conformal
  invariance} (GCI), positing that finite transformations are
well-defined on local fields on Minkowski space itself\footnote{That
  is, our use of the term GCI without further qualification is in
  accord with the statement that a CFT in general satisfies ``GCI on a
  covering space''.}. This assumption not only greatly simplifies the 
analysis by avoiding the difficulties just mentioned. GCI Bose fields
also can have only integer dimensions. They appear as the counterpart 
of chiral fields in two spacetime dimensions, while fields with
anomalous dimensions are expected to arise as intertwiners between
different representations of a local algebra satisfying GCI. 

GCI has far reaching implications, when combined with the Wightman
axioms \cite{SW64}. Together with local commutativity it implies the  
{\it Huygens principle}: the commutator of any two local Bose fields
vanishes for non-isotropic (non-lightlike) separations. Moreover, the
Huygens principle is equivalent (assuming Wightman axioms) to 
{\em strong locality} stating that the commutator vanishes if
multiplied by a sufficiently large power of the Lorentz distance 
square of the fields' arguments. The Huygens principle and energy
positivity yield rationality of correlation functions (\cite{NT01},
Theorem 3.1), as well as meromorphy of (analytically continued)
products of fields acting on the vacuum (\cite[Theorem 9.2]{N05}). 
These results allow one to extend a quantum field theory to 
{\em compactified} Minkowski space $\overline \MM$. 
 
The compactified Minkowski space $\overline\MM$ admits a convenient
complex variable parametrization (\cite{T86})\footnote{For 
  the generalization to arbitrary space-time dimension $D$ and further
  developments see, e.g., \cite{NT04}.}:  
\bea
\label{eq1.1}
\overline\MM = \left\{ z = (z_1,z_2,z_3,z_4) \in
  \CC^4 : \ z = \frac{\bar z}{\bar z^2}\right\}
=\left\{z=e^{i\zeta u},\; \zeta\in\RR, u\in \SS^3\right\}
\qquad (z^2 :=
  \sum_{\alpha = 1}^4 z_{\alpha}^2 = {\mathbf z}^2 + z_4^2) 
\eea
(isomorphic to the $U(2)$ group manifold -- {\it cf.} \cite{U63}). The
embedding $\MM \hookrightarrow \overline\MM$ amounts to a complex conformal
transformation from Minkowski space coordinates $x^{\mu}$, $\mu =
0,1,2,3$ to $z_{\alpha}$: 
$${\mathbf z} = \frac{{\mathbf x}}{\omega (x)} \, , \qquad z_4 =
\frac{1-x^2}{2 \, \omega (x)} \, , \qquad 2 \, \omega (x) = 1 + x^2 - 2
\, i \, x^0 \qquad (x^2 := {\mathbf x}^2 - (x^0)^2) \, . $$
Applying this conformal transformation, we obtain the so called
{\it analytic} ($z$-{\it picture}) in which the correlation functions look
essentially the same as in the $x$-picture. The free massless
Minkowski space 2-point function $(2\pi)^{-2} \, ({\mathbf x}_{12}^2 -
(x_{12}^0 - i \, \varepsilon)^2)^{-1}$ is turned into the distribution 
$\bra\varphi(z_1)\varphi(z_2)\ket = (z_{12}^2)_+^{-1}$ which is
defined as the Taylor series in $z_2$ 
with coefficients in the polynomial algebra $\CC \left[ z_1 ,
  \frac{1}{z_1^2} \right]$. The same applies to the 2-point functions
$\sim (z_{12}^2)_+^{-d}$ of scalar fields $\phi$ of dimension $d$.  

The $z$-picture series expansions provide a convenient description of
the fields as formal power series in $z$ and negative powers of $z^2$
that is completely equivalent to the Wightman approach using
distributions. Together with strong locality, this yields a concept of
higher dimensional vertex algebras introduced in \cite{N05},
generalizing the chiral vertex algebra formalism of \cite{K96}. 

For our purposes it is sufficient to treat the correlators as rational
functions (rather than distributions) and we shall omit the subscript $+$. 

\smallskip

Let $\phi (z)$ be a hermitean scalar field on $\overline\MM$ of dimension 
$d \in \NN$. The OPE of the product of two fields can be written down
as a series of ``bilocal fields'' $V_{\kappa} (z_1 , z_2)$
(\cite{NST02,NST03,NST04,NT04}) with $d-1$ singular terms: 
\bea
\label{eq1.2}
\phi (z_1) \, \phi (z_2) &= &B_{\phi} (z_{12}^2)^{-d} \left\{ 1 + \sum_{\kappa = 1}^{\infty} (z_{12}^2)^{\kappa} \, V_{\kappa} (z_1 , z_2) \right\}  \\
&= &B_{\phi} (z_{12}^2)^{-d} \left\{ 1 + \sum_{\kappa = 1}^{d-1}
  (z_{12}^2)^{\kappa} \, V_{\kappa} (z_1 , z_2) \right\} + \wick{ \phi
(z_1) \, \phi (z_2) } \nonumber
\eea
Here the {\it normal ordered product} $\wick{ \phi (z_1) \, \phi (z_2) }
= \wick{ \phi (z_2) \, \phi (z_1) }$ is defined as the regular part of
the sum in the first line of (\ref{eq1.2}); in particular, it has a local
field limit for $z_{12} \equiv z_1 - z_2 \to 0$. 

Each bilocal field $V_{\kappa}(z_1 , z_2)$ is defined in terms of its
expansion in local symmetric traceless tensors $O_{2\kappa , L} (z)$
($L$ even for $V_{\kappa} (z_1 , z_2) = V_{\kappa} (z_2 , z_1)$) of
fixed {\it twist} ($=$ dimension minus rank) $2\kappa$ -- {\it i.e.}\
of scale dimension $2 \kappa + L$. This implies the orthogonality relation 
\bea
\label{eq1.3}
\bra V_{\kappa} (z_1 , z_2) \, V_{\lambda} (z_3 , z_4) \ket = 0 \quad \hbox{for} \ \kappa \ne \lambda \, .
\eea
The twist two field, $V_1 (z_1 , z_2)$, involves in its expansion an
infinite series of conserved tensor fields, including the
stress-energy tensor $T_{\alpha\beta} (z)$, and is harmonic in each
argument: 
\bea
\label{eq1.4}
\partial^2_{z_1} \, V_1 (z_1 , z_2) = 0 = \partial^2_{z_2} \, V_1 (z_1 ,
z_2), \qquad \partial^2_z := \sum_{\alpha = 1}^4
\frac{\partial^2}{\partial z_{\alpha}^2} \, . 
\eea
For $d=2$ the sum of singular terms in (\ref{eq1.2}) reduces to a
single one, $V_1$, and it was proven (see Sect.\ \ref{sec5}) that
$V_1$ (and $\phi$) can be expressed in this case as a sum of normal
products of (mutually commuting) free massless scalar fields.  

It was then natural to turn attention to a GCI field $\LL(z)$ of
dimension $d=4$ with the properties of a (gauge invariant) Lagrangian
density \cite{NST03,NT04}. Indeed, it is the lowest
dimensional scalar field of $d > 2$ which can be interpreted as a
local observable in a gauge field theory. It involves higher twist
bilocal fields in the expansion (\ref{eq1.2}) which do not obey the
free field equation (\ref{eq1.4}) and thus offer a chance for a
nontrivial GCI theory of local observables. 

\smallskip

The general GCI Wightman 4-point function $\langle 1234\rangle \equiv
\bra \LL  (z_1) \, \LL  (z_2) $ $\LL 
(z_3)\,\LL  (z_4) \ket$ has the form
\cite{NST03,NST04} 
\bea
\label{eq1.5}
\langle 1234\rangle &= & \langle 12 \rangle \langle 34 \rangle +
\langle 13 \rangle \langle 24 \rangle + \langle 14 \rangle \langle 23
\rangle + \frac{B_\LL^2 (z_{13}^2 \, z_{24}^2)^2}{(z_{12}^2 \, z_{23}^2 \,
  z_{34}^2 \, z_{14}^2)^3} \, P_5 (s,t) \nonumber \\ 
&= &
\langle 12\rangle\langle 34\rangle \left\{ 1 + s^4 + s^4\,t^{-4} 
+ \frac{s}{t^3} \, P_5 (s,t) \right\} 
\eea
where $\langle ij \rangle$ stands for the 2-point function $\bra \LL
(z_i) \, \LL (z_j) \ket = B_\LL\;(z_{ij}^2)^{-4}$ ($B_\LL > 0$) and $P_5
(s,t)$ is a crossing symmetric polynomial of overall degree five in
the conformally invariant cross ratios 
\bea
\label{eq1.6}
s = \frac{z_{12}^2 \, z_{34}^2}{z_{13}^2 \, z_{24}^2} \, , \qquad t =
\frac{z_{14}^2 \, z_{23}^2}{z_{13}^2 \, z_{24}^2}\,. 
\eea
The crossing symmetry ($s_{12} \,
  P_5 (s,t) := t^5 \, P_5 \left(\frac{s}{t} , \frac{1}{t} \right) =
  P_5 (s,t)$, $s_{13} \, P_5 (s,t) := P_5 (t,s) = P_5 (s,t)$)
reflects the local commutativity of the field ${\cal L}$. 

The polynomial $P_5$ involves five free parameters and will be
displayed in Section~\ref{sec2}. (If we add to these the normalization
$B^2$ of the product of 2-point functions and view two models
differing by rescaling $\LL  \to \lambda \, \LL$ of the basic
field as equivalent we will end up with a five-dimensional projective
space of 4-point functions described in terms of six homogeneous
coordinates.) 

\smallskip

We study in Section~\ref{sec2} the partial wave expansion of the
4-point function (\ref{eq1.5}) of the scalar field $\LL$,
i.e., the decomposition
\bea
\label{eq1.7}
\langle 1234\rangle =\sum_{\kappa,L} \langle 1234\rangle_{\kappa L}
\equiv \sum_{\kappa,L}  \bra \LL  (z_1) \, {\mathcal
  L} (z_2) \; \Pi_{\kappa L}\; \LL  (z_3) \LL  (z_4)
\ket 
\eea
where $\Pi_{\kappa L}$ is the projection onto the irreducible positive-energy
representation of the conformal group $SU(2,2)$ with $U(1)\times
SU(2)\times SU(2)$ weight $(2\kappa +L,\frac12 L,\frac12 L)$, i.e.,
the representation by a rank $L$ symmetric traceless tensor of scaling
dimension $2 \kappa + L$. Except for the vacuum contribution ($\kappa
= 0$, $L=0$) only positive even twists $2\kappa$ and nonnegative even
spins $L$ appear in the expansion of the product of two identical
neutral scalar fields \cite{NST03}. 

\smallskip

Each partial wave has the form
\bea
\label{eq1.8}
\langle 1234\rangle_{\kappa L} = \langle 12\rangle
\langle34\rangle \cdot
B_{\kappa L} \cdot \beta_{\kappa L} (s,t)
\eea
where $\beta_{\kappa L} (s,t)$ are universal functions kinematically
determined by the representation $\left( 2\kappa + L , \, \frac{1}{2}
  \, L , \, \frac{1}{2} \, L \right)$. Only the coefficients 
$B_{\kappa L}$ are specific for the field $\LL$ and characterize its
couplings to twist $2\kappa$ spin $L$ fields present in the theory. We
compute all partial wave coefficients $B_{\kappa L}$ of the
4-point function (\ref{eq1.5}) in closed form (Appendix~\ref{appB}). 
For the subsequent analysis it is crucial that we can control the
asymptotic behaviour at large $L$, which was not accessible by
previous ``leading order expansions''.

\smallskip

Wightman positivity of the 4-point function is equivalent to the
positivity of the infinite series of coefficients $B_{\kappa L}$
(Section~\ref{sec3}). It turns out that only two of the six structures
entering (\ref{eq1.5}) are separately positive. Nevertheless positivity is
satisfied for the closure of a non-empty open set in the five-dimensional
projective parameter space.  

Of course, the discussion of positivity has to be extended beyond the
4-point level, where we have at this moment only partial control.
For the case of a scalar field of scaling dimension 2, a complete
classification has been obtained \cite{NST02,N-un} by solving a moment  
problem involving the parameters of arbitrary $n$-point functions.  
In the case of two-dimensional CFT, a similar procedure implicitly
extending to higher $n$-point functions by exploiting the
presence of the stress-energy tensor in the operator product expansion
of a neutral field with itself, has been demonstrated to yield the
Friedan-Qiu-Shenker quantization of the central charge and the scaling
dimensions \cite{RS88}.

In the present case, we shall advance the analysis by assuming that
the $d=4$, $L=2$ field appearing in the OPE of two $\LL$'s coincides
with the stress-energy tensor $T_{\alpha\beta}$. This allows us to
relate, in Section~\ref{sec4}, the amplitude $c$ of the (unique
up to a factor) 2-point function of $T$ with three parameters
appearing in the 4-point function (\ref{eq2.1}) of $\LL$. Similar
relations, providing information about the stress-energy tensor from
the OPE of other fields, have been known (and exploited for various
purposes) before, e.g., \cite{P96,AFP00,AEPS02}.

We have to be aware of the possibility of the presence of
free fields in the theory. Especially in the case when the OPE of
$\LL$ with itself contains a scalar field of dimension 2, we know
(\cite{NST02,N-un}, see above) that the subtheory generated by the
latter can be represented by the even Wick polynomials of a massless
free field theory. In this situation, the general structure theory of
subsystems and local extensions (e.g., \cite[Sect.~5]{CC05}) suggests
that the local algebras of the full theory are contained in the tensor
product of the local algebras of the massless free field theory and
those of a decoupled theory
\bea \label{eq1.9}
F(O) \subset A(O) \otimes \widehat F(O)\,. \eea
A field theoretic interpretation of this formula would be a
factorization with the general structure 
\bea \label{eq1.10}
\LL=\wick{\varphi^4} + \varphi \, \widehat X + \widehat\LL 
\eea 
where the scalar fields $\widehat X$ of dimension 3 and $\widehat\LL$
of dimension 4 are local fields associated with the decoupled
theory. In particular, the OPE of $\widehat\LL$ generates no scalar
field of dimension 2. Without loss of generality, one may then assume from
the outset that $\LL$ generates no scalar field of dimension 2, thus
effectively decoupling the free field from the theory.

Unfortunately, neither do we have in our approach sufficient control
over the technical assumptions underlying (\ref{eq1.9}), nor is there
a precise derivation of (\ref{eq1.10}) from (\ref{eq1.9}).
For this reason, we attempt to obtain more direct support for the
structure (\ref{eq1.10}) by studying the field content through
iterated OPE's. This is the motivation for our analysis in
Section~\ref{sec5}, where we draw some first consequences of the
analysis of the 6-point function (sketched in Appendix~\ref{appC})
whose systematic investigation is postponed to future work. 

Our analysis is based to a large extent on a systematic study of
positivity of mixed 4-point functions involving two pairs of scalar fields. 
We demonstrate in particular, that only one among the four possible
6-point function structures of $V_1 (z_1 , z_2)$, that do not vanish
for coinciding arguments, survives. The more interesting (but also
more complicated) four additional 6-point structures will be studied
in a separate publication.  

\smallskip

Free field constructions, which are known to satisfy positivity, only
fill a discrete (measure zero) subset of our positivity domain, thus
leaving room for a nontrivial theory (Appendix~\ref{appA}).  

\section{Partial wave expansion}
\label{sec2}\setcounter{equation}{0} 
The most general 4-point function of a neutral scalar field of scaling
dimension $d=4$ in four space-time dimensions satisfying GCI is
given in the form (\ref{eq1.5}) where the polynomial $P_5$ is
parametrized by 
\bea 
\label{eq2.1}
P_5(s,t)=\sum_{\nu=0}^2a_\nu\cdot J_\nu(s,t) +
  st\;[\,b\cdot D(s,t)+b'\cdot Q(s,t)\,].\eea
Here the three structures $J_\nu(s,t)$ are given by 
\bea
\label{eq2.2}
J_0(s,t)&=& (t^2+t^3) + s^2(1+t^3) + s^3(1+t^2), \nonumber \\
J_1(s,t)&=& (t+t^4) - (t^2+t^3) +s(1+t^4)- 2s(t+t^3)-\nonumber \\ 
&&-s^2(1+t^3)-s^3(1+t^2)-2s^3t+s^4(1+t), \nonumber \\
J_2(s,t)&=& (1+t^5) -2(t+t^4) + (t^2+t^3) - 2s(1+t^4) + s(t+t^3)+\nonumber \\
&& +s^2(1+t^3) + s^3(1+t^2) + s^3t - 2s^4(1+t) + s^5.
\eea
Locality dictates their symmetry under $[s_{13}f](s,t)=f(t,s)$ and
$[s_{23}f](s,t)=s^5f(1/s,t/s)$.  
The origin of these structures \cite{NST03} are the (harmonic) $s$-channel 
twist 2 contributions
\bea
\label{eq2.3}
 \langle 0\vert\; V_1(z_1,z_2)V_1(z_3,z_4)\;\vert 0\rangle = 
\frac{1}{z_{13}^2z_{24}^2}\cdot \sum_{\nu=0}^2 a_\nu\cdot
j_\nu(s,t)
\eea
where $V_1$ is the twist 2 bilocal field introduced in (\ref{eq1.2}), and
\bea
\label{eq2.4} 
j_0&=&t\inv(1+t), \nonumber \\
j_1&=&t^{-2}[(1-t-t^2+t^3)-s(1+t^2)], \nonumber \\
j_2&=&t^{-3}[(1-2t+t^2+t^3-2t^4+t^5)+s(-2+t+t^3-2t^4)+s^2(1+t^3)] ,
\eea
such that $j_\nu/z_{13}^2z_{24}^2$ have pole singularities in
$z_{ij}^2$ ($i=1,2$; $j=3,4$) of degree $\nu+1$. The structures (\ref{eq2.2})
then arise by symmetrization of the resulting contributions to the
full 4-point function.

The last two terms in (\ref{eq2.1}) contribute to twist 4 (and higher)
partial waves: $D$ and $Q$ are degree two symmetric polynomials, $D$
coinciding with the discriminant of the system (\ref{eqB.6}) below:
\bea
\label{eq2.5} 
D(s,t)&=&1 -2t +t^2 -2s(1+t) + s^2, \nonumber \\
Q(s,t)&=&t+s(1+t).
\eea

The partial wave expansion of the 4-point function reads
\bea 
\label{eq2.6} 
\langle 1234\rangle = \langle 12\rangle\langle 34\rangle \cdot
\sum_{\kappa\geq 0,L\geq 0}B_{\kappa L}\cdot\beta_{\kappa L}(s,t).
\eea
where the partial waves $\beta_{\kappa L}$ are fixed kinematically by
the representation theory of the conformal group \cite{DMPPT77}. In contrast,
the coefficients $B_{\kappa L}$ contain dynamical information about
the couplings of the fields in the model. 

The partial wave of the identity operator is $\beta_{00}=1$, hence the
leading term $1$ in (\ref{eq1.5}) corresponds to $B_{00}=1$. The
expansion of the remaining terms is equivalent to  
\bea \label{eq2.7} 
s^4+s^4t^{-4} +st^{-3}P_5(s,t)
=\sum_{\kappa L}B_{\kappa L}\cdot \beta_{\kappa L}(s,t).
\eea

The coefficients $B_{\kappa L}$ of this expansion are explicitly
computed by a method described in Appendix~\ref{appB}, in terms of the
parameters $a_\nu$, $b$ and $b'$ specifying the truncated 4-point
function $P_5(s,t)$, (\ref{eq2.1}). To obtain them, it was essential
to combine the presentation of the partial waves given in closed form
by Dolan and Osborn in terms of ``chiral variables'' \cite{DO01} with
a new expansion formula presented in the Appendix. The result is
that only even $L\geq 0$ occur, as expected, and 
\bea \label{eq2.8} 
\textstyle\frac12\left(2L\atop L\right)B_{1 L}=
a_0+L(L+1)a_1+\frac14(L-1)L(L+1)(L+2)a_2 ,
\eea
while for even $\kappa\geq 2$ (twist 4, 8, \dots) we obtain
\bea \label{eq2.9} 
\textstyle\frac12\left(2\kappa+2L-2\atop\kappa+L-1\right)
\left(2\kappa-4\atop\kappa-2\right)B_{\kappa L} 
&=& (c^4_{\kappa+L}c^3_{\kappa-1}-c^3_{\kappa+L}c^4_{\kappa-1})
+ a_0\cdot(2c^3_{\kappa+L}c^2_{\kappa-1}-2c^2_{\kappa+L}c^3_{\kappa-1}
+2c^3_{\kappa-1}) - \nonumber \\ 
& -&
a_1\cdot(c^3_{\kappa+L}c^2_{\kappa-1}+c^2_{\kappa+L}c^3_{\kappa-1}-c^3_{\kappa-1}-c^3_{\kappa+L})  
+ a_2\cdot c^3_{\kappa+L}c^3_{\kappa-1} \nonumber \\ 
&+& b\cdot(c^2_{\kappa+L}c^2_{\kappa-1}+c^2_{\kappa+L}+c^2_{\kappa-1}-2) 
- b'\cdot (2c^2_{\kappa-1}-1), 
\eea
and for odd $\kappa\geq 3$ (twist 6, 10, \dots) 
\bea \label{eq2.10} 
\textstyle\frac12\left(2\kappa+2L-2\atop\kappa+L-1\right)
\left(2\kappa-4\atop\kappa-2\right)B_{\kappa L}&=& 
(c^4_{\kappa+L}c^3_{\kappa-1}-c^3_{\kappa+L}c^4_{\kappa-1}) 
+ a_0\cdot(2c^3_{\kappa+L}c^2_{\kappa-1}-2c^2_{\kappa+L}c^3_{\kappa-1}
-2c^3_{\kappa+L}) + \nonumber \\ 
&+&a_1\cdot(c^3_{\kappa+L}c^2_{\kappa-1}+c^2_{\kappa+L}c^3_{\kappa-1}-c^3_{\kappa-1}-c^3_{\kappa+L})
- a_2\cdot c^3_{\kappa+L}c^3_{\kappa-1} - \nonumber \\  
&-& b\cdot(c^2_{\kappa+L}c^2_{\kappa-1}+c^2_{\kappa+L}+c^2_{\kappa-1}-2) 
+ b'\cdot (2c^2_{\kappa+L}-1), 
\eea
where $c^p_\nu$ vanish for $\nu<p$, and otherwise
\bea \label{eq2.11} 
c^p_\nu=\frac1{(p-1)!^2}\frac{(\nu+p-2)!}{(\nu-p)!} > 0. 
\eea

Of special interest are the coefficients $B_{10}= 2a_0$ and
$B_{20}=b'$, indicating the presence of a scalar field of dimension 2
and of a scalar field of dimension 4 (e.g., the field $\LL$ itself),
respectively, in the operator product expansion of $\LL$ with itself,
and $B_{12}=\frac13a_0+2a_1+2a_2$ which includes the contribution of
the stress-energy tensor  $\Theta^{\mu\nu}$. As we shall see in Sect.\
4, the value of $B_{12}$ gives a lower bound on the amplitude of the
2-point function of $\Theta^{\mu\nu}$.  

\section{Wightman positivity}\label{sec3}
\setcounter{equation}{0}
We want to study the constraints on the amplitudes, deriving from
Wightman positivity. For the 2-point function, positivity
requires $B_\LL\geq 0$. 
 
For the 4-point function, Wightman positivity has already been
partially exploited by the maximal degree of singularities of the
truncated 4-point function. This excludes partial waves corresponding
to non-unitary representations of the conformal group. To get further
insight, we note that each partial wave (\ref{eq1.8}) amounts to the
insertion of a 
projection $\Pi_{\kappa L}$ into the 4-point function. Because Hilbert
space positivity implies positivity of the quadratic forms $\bra\cdot
\Pi_{\kappa L}\cdot\ket$, each partial wave must separately satisfy
Wightman positivity. Since all $\beta_{\kappa L}$ occur with
positive coefficients in the partial wave expansion of the
4-point function of the massless scalar free field (which is certainly
positive), we know that $\beta_{\kappa L}$ are separately positive. Hence
Wightman positivity of the general 4-point function is equivalent to  
\bea
\label{eq3.1}
 B_{\kappa L}\geq 0\qquad\forall \kappa, L.
\eea

We see from (\ref{eq2.8}--\ref{eq2.10}) that the disconnected part as
well as the structure $J_0$ involve separately only positive partial
wave coefficients, while $J_1$ and $Q$ involve negative coefficients at even
$\kappa$, while $J_2$ and $D$ involve negative coefficients at odd
$\kappa$. To have a positive 4-point function, all negative
contributions of these structures must be dominated by positive
contributions from other structures, giving relative bounds 
between the amplitudes $a_\nu$ and $b$, $b'$.

For $\kappa=1$, (\ref{eq2.8}) must be nonnegative for all $L$:
\bea \label{eq3.2}
a_0+L(L+1)a_1+\textstyle\frac14(L-1)L(L+1)(L+2)a_2\geq 0.
\eea 
At $\kappa=2$, all terms in (\ref{eq2.9}) involving $c^p_{\kappa-1}$ ($p\geq 2$) 
vanish, hence 
\bea \label{eq3.3}
\textstyle \frac14L(L+1)(L+2)(L+3)a_1+L(L+3)b+b'\geq 0.
\eea
For $L=0$ and for $L\to \infty$, these two conditions imply 
\bea \label{eq3.4} 
a_\nu\geq 0 \quad (\nu=0,1,2) \qquad\hbox{and}\qquad b'\geq 0. 
\eea
Note that $2a_0$ and $b'$ are the coefficients $B_{10}$ and $B_{20}$ of
the partial waves due to scalar fields of dimension 2 and 4,
respectively, in the operator product expansion. 

For intermediate $L$, one obtains from $\kappa=2$ a negative lower
bound on $b$ depending on the values of $a_1$ and $b'$:
\bea \label{eq3.5} 
b \geq -\min_{L\,\hbox{\tiny even}}
\left\{\frac{(L+1)(L+2)}4\cdot a_1+\frac1{L(L+3)}\cdot b'\right\}. 
\eea 
If $a_1=0$, then $b\geq 0$. If $b'=0$, then $b\geq -3a_1$. 

Similarly, (\ref{eq2.10}) gives at $\kappa=3$ 
\bea \label{eq3.6}
\textstyle \frac14(L+1)(L+2)(L+3)(L+4)(2a_0+a_1) +
(L+2)(L+3)(2b'-3b)-b' \geq 0.
\eea
These are upper bounds on $b$ of which, in view of (\ref{eq3.4}), $L=0$ is the
strongest one:
\bea \label{eq3.7} 
\textstyle b\leq \frac13(2a_0+a_1)+\frac{11}{18}b'. 
\eea

Further upper bounds arise on $a_1$, $b'$ at even $\kappa\geq 4$, and
on $a_2$, $b$ at odd $\kappa\geq 5$. They are always relative to the
respective remaining amplitudes and to the coefficient $B_{00}=1$ of the
disconnected terms $s^4+(s/t)^4$ in (\ref{eq2.7}). Since at large angular
momentum and large twist, the disconnected contributions dominate all
partial wave coefficients, the strongest bounds will arise from small
values of $L$ and $\kappa$, e.g., 
\bea \label{eq3.8} 
216a_1+11b'&\leq& 120+228a_0+180a_2+88b \qquad (\kappa=4,L=0),
\nonumber \\
900a_2+90b&\leq& 800+980a_0+880a_1+13b' \qquad (\kappa=5,L=0). 
 \eea 

All bounds (\ref{eq3.4}--\ref{eq3.8}) taken together, leave a closed
admitted region in the space of the five amplitudes, with a
nontrivial open interior. Wightman positivity for mixed vectors
spanned by $\LL(\cdot)\Omega$ and $\LL(\cdot)\LL(\cdot)\Omega$ gives
rise to an upper bound on the square of the 3-point amplitude,
relative to the coefficient $B_{20}=b'$, which we don't
display. Clearly, if $B_{20}=0$ then the 3-point function must vanish.

More decisive bounds are expected, however, beyond the 4-point level.
Although these involve more and more new amplitude parameters, it is
expected (e.g., from experience with two-dimensional models) that the
resulting inequalities also put further constraints on the 4-point
amplitudes alone. 

\section{Twist two contribution, the stress-energy tensor and the central charge} 
\label{sec4}\setcounter{equation}{0}

The importance of assuming the existence of a stress-energy tensor
$\Theta_{\mu\nu} (x)$, $\mu , \nu = 0,1,2,3$, in the axiomatic
approach to conformal field theory has been recognized long ago
(\cite{MS72}; see also \cite{M88}). It is a (conserved, symmetric,
traceless rank 2) tensor field of {\it twist 2} (dimension 4), whose
$z$-picture counterpart will be written in the form  
\bea
\label{eq4.1}
T(z,v) := T_{\alpha\beta} (z) \, v^{\alpha} \, v^{\beta} \, , \qquad
\hbox{so that} \quad (\partial_z \cdot \partial_v) \, T (z,v) = 0 =
\partial^2_v \, T (z,v) \, , 
\eea
($\alpha , \beta = 1,2,3,4$); by definition it gives rise to the
generators of infinitesimal conformal transformations. If a field $T$
with such properties exists then it should appear in the OPE of the
product of any local field $\psi$ with its conjugate, $\psi^*$; in
particular, in the product of any hermitean scalar field $\phi$ with
itself. Thus, the partial wave coefficient $B_{12} = \frac{1}{3} \,
a_0 + 2 \, a_1 + 2 \, a_2$ (\ref{eq2.8}) should be strictly positive. 

The normalization of the stress-energy tensor is fixed by the
condition (Ward identity) that its integrals give the generators of
the conformal symmetry. While it is customary to use to this end the
Minkowski energy-momentum components $P_{\mu} = \int \Theta_{\mu 0}
(x) d^3x$, which can be defined as bilinear forms for rational
correlation functions, it is technically simpler to introduce instead
the {\em conformal Hamiltonian} given by the integral over the unit
3-sphere in the $z$-picture  
\bea
\label{eq4.2}
H = \int_{\SS^3} T (u,u)(du)\quad \hbox{for} \ z_{\alpha} = e^{2\pi i
  \zeta} \, u_{\alpha} \, , \qquad (du) = \delta ({u^2} - 1) \,
\frac{d^4 \, u}{2 \, \pi^2} \, . 
\eea

The conformal dimension $d$ of a local field $\phi (z)$ coincides with
the minimal eigenvalue of the conformal Hamiltonian on the
``1-particle space'' ${\rm Span} \, \{ \phi (z) \ket \}$; we have 
\bea
\label{eq4.3}
[H,\phi (z)] = \left( d + z \cdot \partial_z \right)
\phi (z) \qquad \hbox{implying} \quad (H-d) \, \phi (0) \ket = 0 
\eea
(since $H \ket = 0$).

\smallskip

The ratio of the mixed 3-point function to the 2-point function of the
hermitean scalar field $\phi$ of dimension $d$ is uniquely determined
by conformal invariance, the normalization being fixed by
(\ref{eq4.3}) and (\ref{eq4.2}):
\bea
\label{eq4.4}
\bra \phi (z_1) \, T (z_2 , v) \, \phi (z_3) \ket = d \cdot \bra \phi
(z_1) \, \phi (z_3) \ket \, z_{13}^2 \, w_3^{(1)} (z_1 ; z_2 , v ;
z_3) 
\eea
where
\bea
\label{eq4.5}
(z_{12}^2)^d \, \bra \phi (z_1) \, \phi (z_2) \ket = \langle d \mid d
\rangle = B_{\phi} > 0 
\eea
and $w_3^{(1)}$ is the 3-point function $\bra\varphi T\varphi\ket$ of
a (dimension 1) free scalar field $\varphi$ with its stress-energy tensor
(eq.~(\ref{eq4.18})):  
\bea
\label{eq4.6}
w_3^{(1)} (z_1 ; z_2 , v ; z_3) = \frac{2}{3 \, z_{12}^2 \, z_{23}^2}
\left\{ (2 X_{13}^2 \cdot v)^2 - v^2 \, \frac{z_{13}^2}{z_{12}^2 \,
    z_{23}^2} \right\} \, , \qquad X_{13}^2 := \frac{z_{23}}{z_{23}^2} +
\frac{z_{12}}{z_{12}^2} \, . 
\eea
The ket vector $\vert d \rangle$ (and its dual bra vector), appearing
in (\ref{eq4.5}), is defined by the analytic continuation of $\phi (z)
\ket$ ($\bra \phi (z)$): 
$$
\vert d \rangle := \phi (0) \ket \, , \qquad \langle d \vert :=
\lim_{w \to 0} \left\{ (w^2)^{-d} \, \bra \phi \left(\frac{w}{w^2}\right) 
\right\} = \lim_{z \to \infty} \, \{ (z^2)^d \, \bra \phi(z) \} \, , 
$$
so that Eq.~(\ref{eq4.4}) implies the relation
\bea
\label{eq4.7}
\langle d \vert\; T (z,v) \;\vert d \rangle = \frac{2d}{3} \, \langle d
\mid d \rangle \left\{ 4 \, \frac{(z \cdot v)^2}{(z^2)^3} -
  \frac{v^2}{(z^2)^2} \right\} \, . 
\eea
This gives, in accord with (\ref{eq4.2}) and (\ref{eq4.3}),
\bea
\label{eq4.8}
{\textstyle \frac12}\langle d \vert\; T (u,u) \;\vert d \rangle = d \,
\langle d \mid d \rangle = \int_{\SS^3} \langle d \vert\; T (u,u)
\;\vert d \rangle \, (du) = \langle d \vert\; H \;\vert d \rangle \, . 
\eea
Conversely, the 3-point function (\ref{eq4.4}), (\ref{eq4.6}) can be
restored from (\ref{eq4.7}) by a suitable (complex) conformal
transformation. The 3-point function for any other order of factors is
obtained from (\ref{eq4.4}) by using locality. Applying the result to
$\phi = \LL$ ($d=4$) and inserting the expansion (\ref{eq1.2}) for
the product of two $\LL$'s we find 
\bea
\label{eq4.9}
\bra V_1 (z_1 , z_2) \, T (z_3 , v) \ket = 4 \, w_3^{(1)} (z_1 ; z_2 ;
z_3 , v) := \frac 8{3 \, z_{13}^2 \, z_{23}^2}  \, \left\{(2 \,
  X_{12}^3 \cdot v)^2 - v^2 \, \frac{z_{12}^2}{z_{13}^2 \, z_{23}^2}\right\}
\eea
where $X_{12}^3 = \frac{z_{13}}{z_{13}^2} - \frac{z_{23}}{z_{23}^2}$.

\smallskip

We shall now explore the assumption that the traceless
conserved symmetric tensor of dimension 4 appearing in the OPE of 
$\LL (z_1) \, \LL (z_2)$ (or, equivalently in the expansion of $V_1 (z_1 ,
z_2)$) is proportional to the stress energy tensor $T$. (The relation
between the 4-point function of a basic field and the 2-point function
of the stress-energy tensor has been exploited before \cite{P96} in
the context of a conformally invariant $O(N)$ model in $2<D<4$
space-time dimensions.)

The relevant tensor in the OPE is isolated by applying the differential
operator  
\bea
\label{eq4.10}
{\mathcal D}_{12} (v) := {\textstyle\frac{1}{6}} \, \{ (v \cdot \partial_{z_1})^2 + (v
\cdot \partial_{z_2})^2 - 4 \, (v \cdot \partial_{z_1}) \, (v \cdot
\partial_{z_2}) + v^2 \, (\partial_{z_1} \cdot \partial_{z_2}) \} \, 
\eea
to $V_1(z_1,z_2)$ and equating $z_1=z_2$. (The resulting tensor of
dimension 4 is symmetric, traceless and conserved due to the
structure of ${\mathcal D}_{12}$ and due to the fact that
$V_1(z_1,z_2)$ is harmonic.) We thus require that 
\bea 
\label{eq4.11}
T(z,v) = \gamma\cdot {\mathcal D}_{12} (v) V_1(z_1,z_2)\vert_{z_1=z_2=z}
\eea
with some factor of proportionality $\gamma$. We claim that 
$\gamma=\frac{4}{a_0 + 6 \, a_1 + 6 \, a_2}$.
 
Namely, applying the operator ${\mathcal D}_{34} (v)$
to the 4-point function $\bra V_1(z_1,z_2)V_1(z_3,z_4)\ket$ given by 
(\ref{eq2.3}) and
equating $z_3=z_4$ we obtain 
\bea
\label{eq4.12}
{\mathcal D}_{34} (v) \, \{ \bra V_1 (z_1 , z_2) \, V_1 (z_3 , z_4)
\ket \} \vert_{z_4 = z_3} = (a_0 + 6 \, a_1 + 6 \, a_2) \cdot w_3^{(1)}
(z_1 ; z_2 ; z_3 , v) \, , 
\eea
where $w_3^{(1)}$ is defined in (\ref{eq4.9}).
Comparing with the required normalization (\ref{eq4.9}) of $T$, the
factor of proportionality is determined.

Applying also the operator ${\cal D}_{12} (v_1)$ to (\ref{eq4.12}) and
equating $z_1=z_2$, we finally arrive at 
\bea
\label{eq4.13} 
\bra T (z_1 , v_1) \, T (z_2 , v_2) \ket = c
\cdot w_2^{(1)} (z_1 , v_1 ; z_2 , v_2) 
\eea
where
\bea
\label{eq4.14}
w_2^{(1)} (z_1 , v_1 ; z_2 , v_2) = \frac{4}{3 \, (z_{12}^2)^4} \, \{
(2 \, v_1 \, r (z_{12}) \, v_2)^2 - v_1^2 \, v_2^2 \} \, , 
\eea
\bea\label{eq4.15}
r(z) = \un - 2 \, \frac{z \otimes z}{z^2} \qquad \hbox{i.e.} \quad
r_{\alpha\beta} (z) = \delta_{\alpha\beta} - \, 2
\, \frac{z_{\alpha} \, z_{\beta}}{(z^2)_+} \,, 
\eea
and the coefficient is 
\bea \label{eq4.16}
c=\frac{16}{a_0 + 6 \, a_1 + 6 \, a_2}\,.
\eea
In fact, because $T$ is a symmetric traceless and conserved rank 2
tensor field of dimension 4, its 2-point function is necessarily of
the form (\ref{eq4.13}), and the
proportionality coefficient $c$ can be viewed as a
four-dimensional generalization of the central charge\footnote{It
  corresponds to the ``primary central charge'' as discussed in the
  literature -- see \cite{A98} and references therein.} of the
Virasoro algebra.  

In general, when the proportionality (\ref{eq4.11}) fails,
then also the resulting relation (\ref{eq4.16}) between the
amplitude of the 2-point function (\ref{eq4.13}) and the coefficients
of the 4-point function (\ref{eq2.3}) fails. The contribution to the
OPE of $\LL\cdot\LL$ orthogonal to the stress-energy tensor will add
to the partial wave coefficient $B_{12}=(a_0+6a_1+6a_2)/3$, hence one
has in general  
\bea
\label{eq4.17}
a_0+6a_1+6a_2 \geq \frac{16}c \qquad \Leftrightarrow \qquad c\geq
\frac{16}{a_0+6a_1+6a_2}.
\eea
The saturation of this bound, i.e.\ (\ref{eq4.16}), is equivalent to
our stress-energy tensor condition (\ref{eq4.11}). This condition also
excludes the presence of generalized free fields (see
Appendix~\ref{appA}). 

Let us now determine $c$ for the stress-energy tensors of free
fields. For the stress-energy tensor 
\bea
\label{eq4.18} 
T_\varphi (z,v) &:=& D_{12} (v) \; \wick{\varphi (z_1) \, \varphi (z_2)}
\vert_{z_1 = z_2 = z} \nonumber \\ &=& 
\textstyle \frac 13 \; \wick{\left\{ \varphi (z) \, (v \cdot
    \partial_z)^2 \, \varphi (z) - 2 \, [v \cdot \partial_z \, 
    \varphi (z)]^2 + \frac 12 \, v^2\,\partial_z \, \varphi(z) \cdot
    \partial_z \, \varphi (z) \right\}} \, , 
\eea
of a free massless scalar field $\varphi$ (of dimension $d=1$,
$\bra\varphi(z_1)\varphi(z_2)\ket =(z_{12}^2)\inv$), we 
find the 2-point function (\ref{eq4.14}), i.e., $c_\varphi=1$.

For a pair $\psi , \psi^*$ of conjugate (2-component) Weyl spinor
fields of dimension $3/2$ and 2-point function 
\bea
\label{eq4.19}
\bra \psi (z_1) \, \psi^* (z_2) \ket \, ( \equiv \bra \psi (z_1)
\otimes \psi^* (z_2) \ket ) = \frac{2{\not\!z}_{12}^+}{(z_{12}^2)^2} \,
, \ {\not\!v}^+ = \sum_{\alpha = 1}^4 Q_{\alpha}^+ \, v_{\alpha} =
\begin{pmatrix} v_4 + i v_3 &i v_1 + v_2 \\ iv_1 - v_2 &v_4 - iv_3
\end{pmatrix} 
\eea
($Q_j = - \, i \, \sigma_j$ being the imaginary quaternion units,
$Q_j^+ = - \, Q_j$, $j = 1,2,3$) the stress energy tensor $T_{\psi}
(z,v)$ is given by 
\bea
\label{eq4.20}\textstyle
T_{\psi} (z,v) = - (v \cdot \partial_{z_1}) \, (v \cdot
\partial_{z_2}) \, V_1^{\psi} (z_1 , z_2) \vert_{z_1 = z_2 = z} =
\frac{1}{2} \; \wick{ \{ v \cdot \partial_z \, \psi^* (z)) \, {\not\!v} \, \psi
  (z) - \psi^* (z) \, {\not\!v} \, v \cdot \partial_z \, \psi (z) \} } 
\eea
where $V_1^\psi(z_1,z_2)$ is the bilocal twist 2 field in the OPE of
$\psi^*\psi$ with itself, defined as in eq.\ (\ref{eq1.2}),
\bea
\label{eq4.21}\textstyle
V_1^{\psi} (z_1 , z_2) = \frac{1}{4}\; \wick{ (\psi^* (z_1) \,
  {\not\!z}_{12} \, \psi (z_2) - \psi^* (z_2) \, {\not\!z}_{12} \,
  \psi (z_1))} \, . 
\eea
It 
reproduces the term proportional to $j_1 (s,t)$ in the 4-point
function (\ref{eq2.3}) 
\bea
\label{eq4.22}
\bra V_1^{\psi} (z_1, z_2) \, V_1^{\psi} (z_3 , z_4) \ket = 
\frac{j_1 (s,t)}{2 \, z_{13}^2 \, z_{24}^2} \, . 
\eea
Repeated application of (\ref{eq4.20}) gives
\bea
\label{eq4.23}
\bra V_1^{\psi} (z_1 , z_2) \, T_{\psi} (z_3 , v) \ket = 
3 \, w_3^{(1)} (z_1 ; z_2 ; z_3 , v)
\eea
and
\bea
\label{eq4.24}
\bra T_{\psi} (z_1 , v_1) \, T_{\psi} (z_2 , v_2) \ket = 3 \,
w_2^{(1)} (z_1 , v_1 ; z_2 , v_2) \, \qquad \hbox{i.e.}\quad c_\psi=3\,.
\eea
The factor 3 in (\ref{eq4.23}) as compared to 4 in (\ref{eq4.9}) is
correct because $V_1^\psi$ arises from the dimension 3 field $\psi^*\psi$.
The fact that $T_{\psi}$ (\ref{eq4.20}) is properly normalized ({\it
  i.e.}\ satisfies (\ref{eq4.8})) can also be read off the 3-point function
\bea
\label{eq4.25}
\bra \psi (z_1) \, T_{\psi} (z_2 , v) \, \psi^* (z_3) \ket =
\frac{2}{(z_{12}^2 \, z_{23}^2)^2} \, \{ (2v \cdot X_{13}^2) \,
{\not\!z}_{12}^+ \, {\not\!v} \, {\not\!z}_{23}^+ - v^2 \,
{\not\!z}_{13}^+ \} \, . 
\eea

For the free electromagnetic field $F_{\alpha\beta} (z)$ with 2-point function
\bea
\label{eq4.26}
\bra F_{\alpha_1 \beta_1} (z_1) \, F_{\alpha_2 \beta_2} (z_2) \ket =
4(z_{12}^2)^{-2} \{ r_{\alpha_1 \alpha_2} (z_{12}) \, r_{\beta_1
  \beta_2} (z_{12}) - r_{\alpha_1 \beta_2} (z_{12}) \, r_{\beta_1
  \alpha_2} (z_{12}) \} 
\eea
where $r(z)$ is defined in (\ref{eq4.16}) (and satisfies $[r(z)]^2 =
\un$, $r(z) \, z = -z$) and 
\bea
\label{eq4.27}\textstyle \LL ^F (z) = -\frac{1}{4} \; \wick{ F_{\alpha
    \beta} (z) \, F_{\alpha \beta} (z) } 
\eea
we find (according to \cite{NST03}) 
\bea
\label{eq4.28}
\LL ^F (z_1) \, \LL^F (z_2) = \frac{48}{(z_{12}^2)^4} \, \{ 1 +
z_{12}^2 \, V_1^F (z_1 , z_2) + O ((z_{12}^2)^8) \} 
\eea
with
\bea
\label{eq4.29}
V_1^F (z_1 , z_2) &= &\textstyle \frac{1}{24}\, z_{12}^2\cdot
r_{\alpha_1 \alpha_2} (z_{12}) \, r_{\beta_1 \beta_2} (z_{12})\; \wick{
  F_{\alpha_1 \beta_1}   (z_1) \, F_{\alpha_2 \beta_2} (z_2) } =
\nonumber \\ &= &\textstyle\frac{1}{24}\; \wick{ \{ z_{12}^2 \, F_{\sigma \tau} 
(z_1) \, F_{\sigma     \tau} (z_2) + 4 \, z_{12}^{\alpha} \,
F_{\alpha\sigma} (z_1) \, F_{\sigma \beta} (z_2) \, z_{12}^{\beta} \} } \, . 
\eea
This implies
\bea
\label{eq4.30}
\bra V_1^F (z_1 , z_2) \, V_1^F (z_3 , z_4) \ket = \frac{2}{9} \,
\frac{j_2 (s,t)}{z_{13}^2 \, z_{24}^2} \, . 
\eea
Applying (\ref{eq4.12}) to (\ref{eq4.30}) for $a_0 = a_1 = 0$, $a_2 =
\frac{2}{9}$, we find 
\bea
\label{eq4.31}
\bra V_1^F (z_1 , z_2) \, T_F (z_3 , v) \ket = 3 \, {\mathcal D}_{34}
(v) \left\{\frac29 \frac{j_2 (s,t)}{z_{13}^2 \, z_{24}^2} \right\}
\biggl\vert_{z_4 = z_3} = 4 \, w_3^{(1)} (z_1 ; z_2 ; z_3 , v) \, , 
\eea
and hence,
\bea
\label{eq4.32}
\bra T_F (z_1 , v_1) \, T_F (z_2 , v_2) \ket = 3 \, {\mathcal D}_{01}
(v_1) \, \bra V_1^F (z_0 , z_1) \, T_F (z_2 , v_2) \ket \vert_{z_0 =
  z_1} = 12 \, w_2^{(1)} (z_1 , v_1 ; z_2 , v_2) \, . 
\eea

To summarize, the three basic free field theories reproduce
the three structures in the general conformal invariant 3-point
function of $T$ (see \cite{S88}), as well as the three contributions
$j_a (s,t)$, $a = 0,1,2$, to the twist 2 part of the 4-point
function of $\LL$. For the ``central charge'' $c$ defined in terms of
the 2-point function (\ref{eq4.13}) of $T$, we find the following values:  
\bea
\label{eq4.33}
\bra T_X (z_1 , v_1) \, T_X (z_2 , v_2) \ket = c_X \cdot w_2^{(1)} (z_1 ,
v_1 ; z_2 , v_2) \, , \ c_{\varphi} = 1 \, , \ c_{\psi} = 3 \, , \ c_F
= 12\, .
\eea
(For a Dirac spinor, one gets twice the value for the Weyl spinor.)

\medskip

\noindent {\it Remark 4.1:}  In Minkowski space notation the generic
2-point function of the stress energy tensor $\Theta_{\mu\nu}(x)$, the
counterpart of (\ref{eq4.33}), has the form 
$$
\bra \Theta_{\kappa\lambda} (x_1) \, \Theta_{\mu\nu} (x_2) \ket =
\frac{8 \, c}{3 \, (2 \, \pi)^4} \, (x_{12}^2)_+^{-4} \left(
  r_{\kappa\mu} (x_{12}) \, r_{\lambda\nu} (x_{12}) + r_{\kappa\nu}
  (x_{12}) \, r_{\lambda\mu} (x_{12}) - \frac{1}{2} \,
  \eta_{\kappa\lambda} \, \eta_{\mu\nu} \right) 
$$
where
$$
r_{\lambda\mu} (x_{12}) = \eta_{\lambda\mu} - 2 \,
\frac{(x_{12})_{\lambda} \, (x_{12})_{\mu}}{(x_{12}^2)_+} \, , \qquad
(x_{12}^2)_+ = {\mathbf x}_{12}^2 - (x_{12}^0 - i \, \varepsilon)^2 \,. 
$$

\section{Positivity restrictions on the six-point function of $V_1$
  and factoring the $d=2$ contribution} 
\label{sec5}\setcounter{equation}{0}

The positivity conditions for the 4-point functions restrict the
possible structures of the 6-point function. We shall demonstrate this
studying the four structures of
$\bra V_1 (z_1 , z_2) \, V_1 (z_3 ,z_4) \, V_1 (z_5 , z_6) \ket$
displayed in Appendix~\ref{appC} for
which the local field $\phi (z) = \frac{1}{2} \, V_1 (z,z)$ of
dimension 2 has a non-vanishing contribution. Such structures can only
appear if $a_0$ in (\ref{eq2.1}) is non-zero as (noting that $j_{\nu}
(0,1) = 2 \, \delta_{\nu 0}$)
\bea
\label{eq5.1}
\phi(z):=\frac 12 \;V_1 (z,z) \quad \Rightarrow \quad \bra \phi (z_1) \,
\phi (z_2) \ket = \frac{a_0}{2 \, (z_{12}^2)^2} \, .
\eea
The 6-point functions $F_1$, $F_2^{(1)}$, $F_3^{(1)}$ and $F_3^{(2)}$
of Appendix~\ref{appC} contribute to the GCI 4-point function
\bea
\label{eq5.2}
z_{34}^2 \, \bra V_1 (z_1 , z_2) \, \phi (z_3) \, \phi (z_4) \ket =
\frac{1}{z_{13}^2 \, z_{24}^2} \sum_{\nu = 0}^2 A_{\nu} \,
j_{\nu} (s,t) \, = \,
\bra \phi (z_1) \, \phi (z_2) \, V_1 (z_3 , z_4) \ket \, z_{12}^2
\eea
with certain (real) coefficients $A_\nu$.
We conclude from (\ref{eq5.2})
that the 3-point function of $\phi$ is given by
\bea
\label{eq5.3}
\bra \phi (z_1) \, \phi (z_2) \, \phi (z_3) \ket = \frac{A_0}{z_{12}^2
  \, z_{23}^2 \, z_{13}^2} \,. 
\eea

In what follows we shall combine results of earlier work with
implications of Hilbert space positivity on the mixed 4-point
functions like (\ref{eq5.2}), in order to restrict the number and the
values of the parameters involved in the above 6-point function.

\medskip

We first apply results of \cite{NST02} and \cite{N-un} concerning the
subtheory generated by the field $\phi (z)$ which exploit the Wightman
positivity condition for the set of {\it all} $n$-point functions of $\phi$.
One proves that $\phi (z)$ can be presented as a sum
of a generalized free field and Wick squares of independent massless
scalar fields. In the present situation, the assumed presence of a
stress-energy tensor excludes the generalized free field. 

More precisely, one shows (by solving a moments problem \cite{N-un})
that $\phi (z)$ can be decomposed as: 
\bea
\label{eq5.4}
\phi (z) \, = \, \sum_{k \, = \, 1}^N \, \alpha_k \, \phi_k (z)
, \qquad \bra \phi_j (z_1) \, \phi_k (z_2) \ket \, = \,
\frac{c_k}{2} \ \delta_{jk} \, (z_{12}^2)^{-2} ,
\eea
where $\alpha_k$ are {\it all different}, and the absolute
normalizations of the fields $\phi_k$ are fixed by the algebra%
\footnote{This algebra is shown to hold on the vacuum state space of
  the subtheory generated by $\phi$. By the Reeh-Schlieder theorem, it
  holds also on the full state space of the field $\LL$.} 
\bea
\label{eq5.5}
\phi_j (z_1) \, \phi_k (z_2) \, = \,
\delta_{jk} \left(\frac{c_k}{2 (z_{12}^2)^2} + \frac{1}{z_{12}^2} \,
  V_1^{(k)} (z_1,z_2)  \right) 
+ \wick{ \phi_j (z_1) \, \phi_k (z_2) }\;, \qquad 
\phi_k (z) \, = \, \frac{1}{2} \ V_1^{(k)} (z,z) .
\eea
Alternatively, one may first define the amplitudes $c_k$ in a
normalization-independent way by the {\em homogeneous} relation 
$$
c_ k \, ( \bra \phi_k (z_1) \, \phi_k (z_2) \, \phi_k (z_3) \ket )^2
\, = \, 8\bra \phi_k (z_1) \, \phi_k (z_2) \ket \; \bra \phi_k (z_2)
\, \phi_k (z_3) \ket \; \bra \phi_k (z_1) \, \phi_k (z_3) \ket 
$$
and then normalize $\phi_k$ as in (\ref{eq5.5}).

Upon iteration of the OPE, arbitrary linear combinations of the fields
$\phi_k$ can be generated, so the latter belong separately to the
subtheory generated by $\phi$. 

Finally, the amplitudes $c_k$ are proven to be positive integers (see
Theorem 5.1 of \cite{NST02}) and each $\phi_k$ can be represented, at
the expense of extending the state space, as a sum of normal squares
of commuting free massless fields $\{\varphi_{jk} (z) : 1 \leq j \leq c_k\}$,
\bea
\label{eq5.6}
\phi_k (z) \, = \, \frac{1}{2} \sum_{j \, = \, 1}^{c_k}
\wick{\varphi_{jk}^2 (z)}\; , \qquad V_1^{(k)} (z_1,z_2) = 
\wick{\varphi_k (z_1) \, \varphi_k (z_2)}\;.  
\eea

Using eqs.\ (\ref{eq5.1}-\ref{eq5.6}) one finds
\bea
\label{eq5.7}
2 (z_{12}^2)^2 \, \bra \phi (z_1) \, \phi_k (z_2) \ket \, = \, c_k \, \alpha_k
, \qquad a_0 \, = \, \sum_{k \, = \, 1}^N c_k \, \alpha_k^2 \,
, \qquad A_0 \, = \, \sum_{k \, = \, 1}^N c_k \, \alpha_k^3
\eea
and
\bea
\label{eq5.8}
z_{13}^2 \, z_{24}^2 \ \bra V_1^{(j)} (z_1,z_2) \, V_1^{(k)} (z_3,z_4)
\ket \, = \, \delta_{jk} \, c_k \, j_0 (s,t) . 
\eea

The integers $c_k$ also determine the central charge $c$ of (\ref{eq4.13}) of
the stress-energy tensor $T_{\phi} (z,v)$ of the subtheory generated
by the field $\phi$ (or, equivalently, by $\phi_k$, $k = 1,\dots, N$):
\bea
\label{eq5.9}
c_{\phi} \, = \, \sum_{k \, = \, 1}^N c_k .
\eea

Since $V_1 (z_1,z_2)$ and $V_1^{(k)} (z_1,z_2)$ are built only of
twist 2 contributions we have, in view of~(\ref{eq5.5}):
\bea 
\label{eq5.10}
\bra V_1 (z_1 , z_2) \, \phi_j (z_3) \, \phi_k (z_4) \ket
\, = \, \delta_{jk} \, (z_{34}^2)^{-1} \, 
\bra V_1 (z_1 , z_2) \, V_1^{(k)} (z_3 , z_4) \ket .
\eea
By Proposition~4.3 of \cite{NT01} the power of, say, $(z_{23}^2)^{-1}$ 
appearing in the 4-point function $\bra \LL (z_1)$ $\LL (z_2)$ $\phi_k
(z_3)$ $\phi_k (z_4) \ket$ does not exceed two. It follows from
Eq.~(\ref{eq5.10}) and the property that  
the singularities of $\bra V_1 (z_1,z_2)$ $\phi_k (z_3)$ $\phi_k (z_4) \ket$ 
do not exceed those of
$\bra \LL (z_1) \, \LL (z_2) \, \phi_k (z_3) \, \phi_k (z_4) \ket$, that 
$A_2$ in (\ref{eq5.2}) {\it vanishes}, and that we can write
\bea 
\label{eq5.11}
\bra V_1 (z_1 , z_2) \, V_1^{(k)} (z_3 , z_4) \ket
\, = \frac{c_k \, \alpha_k \, j_0 (s,t) + c_k \, A_{1k} \, j_1
  (s,t)}{z_{13}^2 \, z_{24}^2} \, =  \bra V_1^{(k)} (z_1 , z_2) \, V_1 (z_3, z_4) \ket 
\eea
for some real constants $A_{1k}$ ($k = 1,\dots, N$). The coefficients
in front of $j_0 (s,t)$ in (\ref{eq5.11}) are fixed by (\ref{eq5.1})
and (\ref{eq5.4}) and (\ref{eq5.7}) in the limits $z_1 \to z_2$ and
$z_3 \to z_4$. The constants $A_{1k}$ are related to $A_1$ of (\ref{eq5.2}) by 
\bea 
\label{eq5.12}
A_1 \, = \, \sum_{k \, = \, 1}^N
c_k \, \alpha_k^2 \, A_{1k} .
\eea

On the other hand, we observe that for a general bilocal field $W
(z_1,z_2)$ which is conformal invariant of weight $(\kappa,\kappa)$
the field 
\bea 
\label{eq5.13} \textstyle 
C_{12}^{(\kappa)} \, W (z_1 ,z_2) \, := \,
\left\{\frac{1}{2} z_{12}^2\,
  r_{\alpha\beta}(z_{12})\partial_{z_1\alpha}\partial_{z_2\beta} +
  \kappa \, z_{12} \cdot (\partial_{z_1} - \partial_{z_2}) \right\} W (z_1,z_2)
\eea
is again a  bilocal field of weight $(\kappa,\kappa)$.
This follows by the fact that $C_{12}^{(\kappa)}$ is constructed as a
conformal (2-point) Casimir operator on the space of weight
$(\kappa,\kappa)$ bilocal fields. In particular, for $\kappa = 1$, the
conformal partial waves $\beta_{1L}$ are eigenfunctions 
of $C_{12} :=  z_{13}^2 z_{24}^2 \circ C_{12}^{(1)} \circ
(z_{13}^2 z_{24}^2)\inv$ of eigenvalue $L(L+1)$: 
\bea 
\label{eq5.14}
\left( C_{12} - \lambda \right) \beta_{1L} (s,t) \, = \, 0
, \quad \lambda := L (L+1) \,.
\eea
(The operator $C_{12}$ has been introduced in \cite{DO04} in order to
reduce the problem of finding partial waves to an eigenvalue problem.) 
Accordingly, as one verifies directly,
\bea 
\label{eq5.15}
C_{12}^{(1)} \ \frac{j_0 (s,t)}{z_{13}^2 \, z_{24}^2} \, = \,
\frac{j_1 (s,t)}{z_{13}^2 \, z_{24}^2} 
, \qquad C_{12}^{(1)} \ \frac{j_1 (s,t)}{z_{13}^2 \, z_{24}^2} \, = \,
2 \, \frac{j_1 (s,t) + 2 j_2 (s,t)}{z_{13}^2 \, z_{24}^2} .
\eea
Therefore, if we introduce the bilocal field $V_1^{\perp} (z_1,z_2)$ by
\bea
\label{eq5.16}
V_1 (z_1,z_2) =
V_1^{\perp} (z_1,z_2) \, + \, \sum_{k \, = \, 1}^N
\left( \alpha_k \, V_1^{(k)} (z_1,z_2) \, + \, A_{1k} \, C_{12}^{(1)}
  \, V_1^{(k)} (z_1,z_2) \right) 
\eea
then it follows that
\bea
\label{eq5.17}
\bra V_1^{\perp} (z_1,z_2) \, V_1^{(k)} (z_3,z_4) \ket \, = \, 0 .
\eea

We summarize this discussion in 

\medskip 

\noindent {\bf Proposition 5.1: 
\it The bilocal field $V_1^{\perp} (z_1,z_2)$
is orthogonal to any product of $\phi$'s:
\bea
\label{eq5.18}
\bra V_1^{\perp} (z_1,z_2) \, \phi (z_3) \, \dots \, \phi (z_n)  \ket \, = 
\, 0 \qquad (n = 3,4, \dots).
\eea
In other words, the bilinear form defined by $V_1^\perp$ vanishes on
the cyclic subspace $\Pi_{\phi}\HH$ generated by polynomials in $\phi$
acting on the vacuum:
\bea
\label{eq5.19}
\Pi_{\phi} \, V_1^{\perp} (z_1,z_2) \, \Pi_{\phi} \, = \, 0 .
\eea}


\noindent
{\it Proof.} $\Pi_{\phi}$ is the projection on the vertex subalgebra
generated by $\phi (z)$: working in this purely algebraic approach we
do not encounter any domain problem.
Then we note that $\Pi_{\phi} \, V_1^{\perp} (z_1,z_2) \, \Pi_{\phi}$ is
strongly bilocal with respect to $\phi (z)$.
Thus, if we prove that it vanishes acting on the vacuum, then (\ref{eq5.19})
will follow from the Reeh-Schlieder theorem. To this end we observe
that $\Pi_{\phi} \, V_1^{\perp} (z_1,z_2) \, \Omega$ is a vector
containing only twist 2 contributions so that it is contained in the
linear span of the 2-particle spaces $V_1^{(k)} (z_1,z_2) \, \Omega$.
But $V_1^{\perp} (z_1,z_2) \, \Omega$ has zero projection on the latter space
because of~(\ref{eq5.17}),
which completes the proof.

\medskip
We interpret Proposition 5.1 as a ``decoupling'' of the subtheory of 
massless free fields or their Wick products. The presence of the free
bilocal fields $V_1^{(k)}$ in the OPE of $\LL$ with itself suggests
the presence of the free massless scalar fields $\varphi$ as factors
in $\LL$, while the orthogonal contribution $V_1^{\perp}$ is due to
factors which decouple from the free scalars. This indicates a structure
of the general form (\ref{eq1.10})
\bea 
\label{eq5.20}
\LL=\wick{\varphi^4} + \varphi \widehat X + \widehat\LL 
\eea 
where $\widehat X$ and $\widehat\LL$ are scalar fields whose OPE
generates no scalar of dimension 2. $\widehat \LL$ would then be the
candidate for a nontrivial field of dimension 4. Since 
Prop.\ 5.1 does not suffice to draw such a conclusion, we try to 
get further insight into the structure of $\LL$ by considering its 
mixed 4-point
functions $\bra \phi_i(z_1)\,\LL(z_2)\,\LL(z_3)\,\phi_i(z_4)\ket$ 
and the OPE
\bea \label{eq5.21}
\phi(z_1)\LL(z_2) = \frac 1{(z_{12}^2)^2}\left\{\widetilde\phi(z_2) +
  V_{02}(z_1,z_2) + O(z_{12}^2)\right\} 
\eea
where $\phi$ is any real scalar field of dimension 2 (which can be
either one of the basic fields $\phi_i$ or some linear combination
such as (\ref{eq5.4})). In particular, we wish to derive and exploit
further implications of Wightman positivity for these 4-point functions.

The local field in (\ref{eq5.21})
\bea \label{eq5.22}
\widetilde\phi(z) = \lim_{\varepsilon\to 0}
\left\{(\varepsilon^2)^2\phi(z+\varepsilon)\LL(z)\right\} 
\eea
is another scalar field of dimension 2, hence again a linear combination of
the $\phi_i$ (which may vanish). $V_{02}$ is a hypothetical bilocal
field of dimension $(0,2)$ with the properties that $V_{02}(z,z)=0$
but $V_{02}(z_1,z_2)\neq 0$ for $z_{12}^2=0$, $z_{12}\neq0$. We shall
now show that such a field must vanish. 

The expansion of $V_{02}$ in local fields involves only {\em twist 2}
tensor fields $O_L(z;z_{12}) := 
O_{\alpha_1\cdots\alpha_L}(z)z_{12}^{\alpha_1} \cdots z_{12}^{\alpha_L}$ 
with $L>0$. The 2-point function of $O_L$ satisfies the conservation
law $\partial_z\cdot\partial_v \bra O_L(z;v) O_L(z_2;v_2)\ket =0$,
hence by Wightman positivity and the Reeh-Schlieder theorem, the
conservation law holds for the field itself. On the other hand, the
unique conformally invariant 3-point function $\bra A(z_1) B(z_2)
O_L(z_3;v) \ket$ satisfies the conservation law only if the scalar fields 
$A$ and $B$ have equal dimension. Therefore, $O_L$ cannot contribute to
the OPE (\ref{eq5.21}) of $\phi$ with $\LL$.

More generally, the above argument allows to prove the following
complement to Proposition 4.3 of \cite{NT01}. 

\medskip

\noindent {\bf Proposition 5.2: 
\it Let $A(z)$ and $B(z)$ be two scalar fields of different (positive
integer) dimensions $d_A$ and $d_B$ such that $d_A+d_B=:2m$ is
even. Then the bilocal twist 2 field $V_{AB}(z_1,z_2)$ defined on the
light-like surface $z_{12}^2=0$ by $V_{AB}(z_1,z_2) =
\lim_{z_{12}^2\to0} \{(z_{12}^2)^{m-1} A(z_1)B(z_2)\}$ only involves a
scalar field of dimension 2 in its expansion in local fields. } 

\medskip

In the case at hand, we have $m=3$ and $V_{\phi\LL}(z_1,z_2) =
\widetilde\phi(z_2)$. 
Applying this result to the 4-point function of $\phi_i$ and $\LL$, we find
that $A_{1i}=0=A_1$ (in particular, the subtraction involving the
Casimir operator in (\ref{eq5.16}) is absent), and 
\bea \label{eq5.23} 
&& \bra \phi_i(z_1)\LL(z_2)\LL(z_3)\phi_i(z_4)\ket =
B_\LL\left\{\frac{c_i}{2(z_{14}^2)^2(z_{23}^2)^4} + \right.
\nonumber \\ 
&&+ 
\left.\frac{\tilde b_i}{(z_{23}^2)^2}
\left(\frac 1{(z_{12}^2z_{34}^2)^2}+(2\leftrightarrow 3)\right) + 
\frac{\tilde c_i}{z_{12}^2z_{13}^2(z_{23}^2)^2z_{24}^2z_{34}^2} + 
\frac{\alpha_ic_i}{z_{14}^2(z_{23}^2)^3} 
\left(\frac 1{z_{12}^2z_{34}^2}+(2\leftrightarrow 3)
\right)\right\} \nonumber \\
&& \qquad\qquad= \frac {B_\LL}{(z_{12}^2)^2(z_{23}^2)^2(z_{34}^2)^2}
\left\{\frac{c_i}2  \; s^2/t^2 + \tilde b_i\;(1+s^2) + \tilde c_i \; s 
+ \alpha_ic_i\; (s/t+s^2/t) \right\} \qquad \eea
with new undetermined amplitudes $\tilde b_i$, $\tilde c_i$, while the
other amplitudes have been determined using eqs.\ (\ref{eq5.4}),
(\ref{eq5.5}) and (\ref{eq5.11}). Here, the leading constant term in
the braces (with coefficient $\tilde b_i$) represents the 
{\em complete} twist 2 partial wave $\beta^{\delta=2}_{10} = 1$ (see
eq.\ (\ref{eqB.8})).   
The next-to-leading terms $s$ and $s/t$ can be expanded into partial waves
of twist 4 while all other terms are of higher twist. Performing the
exact partial wave expansion by the method explained in
Appendix~\ref{appB}, one finds 
\bea \textstyle B_{1L} & = &\delta_{L0}\,\tilde b_i, \nonumber \\ 
\textstyle \left({2L+2}\atop L \right ) B_{2L} & = &
\textstyle  \alpha_i c_i\cdot \left({(L+1)(L+2)}\atop 2\right) +
\tilde c_i\cdot  (-1)^L , \\
\textstyle \left({2\kappa+2L-2}\atop{\kappa+L}\right)
\left({2\kappa-4}\atop{\kappa-1}\right) B_{\kappa L} & = & 
\textstyle 2c_i\cdot\left(\left({\kappa-1}\atop 2\right)\left({\kappa+L+1}\atop
    4\right)
-\left({\kappa}\atop 4\right)\left({\kappa+L}\atop 2\right)\right) -
\qquad\qquad\qquad
\nonumber \\ \textstyle - \alpha_ic_i\cdot (-1)^\kappa \left(\left({\kappa+L}\atop
    2\right) + (-1)^L\left({\kappa-1}\atop 2\right)\right) & + &  \tilde
b_i\cdot (-1)^L\left((\kappa+L+1)(\kappa+L-2)-\kappa(\kappa-3)\right)
\qquad (\kappa\geq 3). \nonumber
\eea
From the twist 2 and twist 4 coefficients we see that Wightman positivity
requires $\tilde b_i$ and $\alpha_i c_i$ to be nonnegative, while
we already know that $c_i$ are positive and $\alpha_i\neq 0$. Hence we
conclude   
\bea \label{eq5.24} 
\tilde b_i \;(\equiv B_{10}) \geq 0,\qquad \alpha_i > 0 \quad
\hbox{for}\quad i=1,\ldots N. 
\eea
It then follows that $A_0$ in (\ref{eq5.2}) and (\ref{eq5.3}) is also positive.
Further nontrivial bounds among the amplitudes arise at small values
$\kappa=2,3,4$, $L=0,1$:
\bea 
\textstyle -\alpha_i\,c_i\leq \tilde c_i\leq3\alpha_i\,c_i,
\qquad\qquad (-3+ \frac32\alpha_i)\,c_i\leq\tilde b_i \leq 
(1+\frac12\alpha_i) \,c_i, \qquad\qquad \alpha_i\leq 4 \,. 
\eea
When $\alpha_i$ takes the minimal value 0, then $\tilde c_i=0$ is
fixed; when $\alpha_i$ takes the maximal value 4, then $\tilde b_i=3$
is fixed. All these bounds are clearly consistent with the structure 
(\ref{eq5.20}), and we expect that considerations of this kind should
ultimately allow to prove it.

Assuming for the moment (\ref{eq5.20}) to be correct, and turning to
$\widehat\LL$ instead of $\LL$, we may return to our discussion of
Section~\ref{sec3} and assume $a_0=0$ in (\ref{eq2.1}), which means
that a scalar field of dimension 2 does {\em not} occur in the OPE of  
$\LL$ with itself. This strengthens the bounds on the other
parameters. Let us assume in addition the absence of a dimension
4 scalar in the OPE, i.e., $b'=0$ (in particular, the 3-point function
of $\LL$ vanishes, which is a characteristic property for the
Lagrangian of a pure gauge field \cite{NST03}). Then (\ref{eq3.5}) and
(\ref{eq3.6}) reduce to the positivity restrictions for the remaining
parameters  
\bea
\label{eq5.25}
a_1 \geq 0 \, , \ a_2 \geq 0 \quad (a_1 + a_2 > 0), \qquad -3 \, a_1
\leq b \leq \frac{1}{3} \, a_1 \, .
\eea
The case $a_1 = 0$ (hence $b=0$) is expected to reduce to a direct 
sum of free Maxwell fields. It is thus interesting to study implications 
of positivity at 6- (and higher) point level for $a_1 > 0$.

\section{Conclusion}
\label{sec6}\setcounter{equation}{0}
Our central result is an exhaustive study of the consequences of 
Wightman positivity at the 4-point level, summarized in Sect.\ 3, of 
a GCI scalar field $\LL$ of scaling dimension 4 in four space-time
dimensions. It is based on an explicit computation of all partial wave
coefficients $B_{\kappa L}$, presented in Appendix~\ref{appB}. 
It turns out that the behaviour at asymptotically large spin $L$ gives
rise to non-trivial constraints, which were not accessible with recursive
techniques yielding only the coefficients of a finite number of
``leading'' partial waves. 

We analyze in Sect.\ 4 the consequences of the existence and 
uniqueness of a stress-energy tensor $T$ whose appropriate moments 
generate the infinitesimal conformal transformations. Defining the
{\em central charge} $c$ by the normalization of the 2-point function
(\ref{eq4.13}) of $T$, we establish a lower bound (\ref{eq4.17}) for $c$ 
in terms of the parameters of the twist 2 contributions to the 4-point
function of $\LL$. This bound is saturated if the stress-energy tensor
is precisely the conserved rank 2 tensor appearing in the OPE of the
bilocal field $V_1$ defined by (\ref{eq1.2}) and (\ref{eq1.4}).

Section~\ref{sec5} exploits the results of previous work to analyze
the consequences of the presence of a non-zero scalar field
$\phi(z)=\frac12 V_1(z,z)$ of dimension 2. We prove in this case that
$V_1$ admits an orthogonal decomposition of the form
\bea 
V_1 = V_1^{(\phi)} + V_1^\perp
\eea
where $V_1^{(\phi)}$ belongs to the algebra generated by the scalar
dimension 2 field $\phi$, and $V_1^\perp$ has vanishing correlation
functions with any number of fields $\phi$. We interpret this
decomposition, together with further results in Section~\ref{sec5},
as a strong evidence in favour of a factorization of the theory
according to (\ref{eq1.9}), (\ref{eq1.10}), allowing one to split off
the free field content. 

The positivity condition for the 4-point function 
$\bra\phi(z_1)\LL(z_2)\LL(z_3)\phi(z_4)\ket$ strongly restricts the
parameters of the 6-point function of $V_1$ (discussed in
Appendix~\ref{appC}), and of its 4-point function. 

These results are the first steps towards a more systematic
analysis of Wightman positivity at the 6-point level, and to the 
identification of a nontrivial field of dimension 4.

\vskip10mm

\centerline{\bf \large Acknowledgments}

\medskip
I.T. would like to thank Yassen Stanev, and K.-H.R. would like to thank
Sebastiano Carpi for helpful discussions at an early stage of this work.  
A major part of this work was done while N.N. and I.T. were visiting the
Institut f\"ur Theoretische Physik der Universit\"at G\"ottingen as 
an Alexander von Humboldt research fellow and a recipient of an AvH
Research Award, respectively. A first draft was written (and
corrected) during the stay of I.T. at l'Institut des Hautes \'Etudes
Scientifiques, Bures-sur-Yvette (and at the Theory division of CERN). The 
authors thank these institutions for hospitality and support. 
The work of N.N. and I.T. was supported in part by the Research
Training Network within Framework Programme 5 of the European
Commission under contract HPRN-CT-2002-00325 and by the Bulgarian
National Council for Scientific Research under contract PH-1406. 

\bigskip

\begin{center} \Large\bf Appendix\end{center}

\appendix

\section{Free field constructions}\label{appA}
\setcounter{equation}{0}
In order to see which of the possible values of the parameters in
(\ref{eq2.1}) are with certainty realized in Wightman positive theories, 
we have considered various scalar fields $\LL_I$ of dimension 4 which
can be constructed as Wick products of free fields. These models give
us admissible points $\bfa_I$ in the five-dimensional space of
relative amplitudes $\bfa \equiv(a_0,a_1,a_2,b,b')$ (i.e., ignoring
the absolute normalizations $B_\LL$). 

Before we display these values below, we observe that by taking the
sum $\LL=\LL_1+\LL_2$ of any two commuting fields (defined on the
tensor product of their individual Hilbert spaces $\HH_1$ and $\HH_2$),
the absolute amplitudes $B_\LL$ and $B_\LL\cdot\bfa$ are additive, so
that the vector $\bfa$ for the sum equals $\lambda^2 \bfa_1 +
(1-\lambda)^2\bfa_2$ where $\lambda = B_1/(B_1+B_2)$. For 
sums of more than two commuting fields, one finds the general
``subconvex'' composition law 
\bea
\bfa = \sum \lambda_I^2\bfa_I \qquad\hbox{with}\quad\lambda_I\geq0,\;
\sum\lambda_I=1.
\eea
By taking subconvex combinations of the points $\bfa_I$ realized by
canonical free field models, one obtains a subset with a non-empty open
interior in the five-dimensional parameter space defined by
(\ref{eq3.1}).
In particular, taking sums of copies of identical fields $\LL$, one
can obtain any multiple $\mu\cdot \bfa$ with $0\leq \mu\leq 1$.
We observe, however, that none of the free field models has $b<0$,
although 4-point positivity does not exclude this possibility.

Note also, that the assumption (\ref{eq4.11}) of uniqueness of the
stress-energy tensor introduced in Sect.\ 4 (requiring the $\kappa=1,L=2$
contribution to the operator product expansion to be a multiple of the
stress-energy tensor, and consequently $c=\frac{16}{a_0+6a_1+6a_2}$)
is not stable under taking arbitrary weighted sums, because the OPE
contains the stress-energy tensors $T_I$ with relative weight
$\lambda_I$. This assumption is therefore consistent only with a
parameter subset of measure zero.

\subsection{Canonical free fields}
The scalar free field $\varphi$ in four space-time dimensions has
dimension 1, so one obvious candidate is $\LL=\wick{\varphi^4}$. More
generally, scalar fields of dimension 4 can be obtained as Wick
products of any four components $\varphi_a$ of a scalar free field
multiplet. 

The canonical vector field has dimension 3, so there is no way of
constructing a scalar of dimension 4 with it. 

The canonical Dirac spinor has dimension $\frac 32$. With
this, one can construct the scalar Yukawa-type Wick product
$\varphi\;\wick{\bar\psi\psi}$ of dimension 4.   

Finally, there is the gauge model $\LL=\frac14\,\wick{F_{\mu\nu}F^{\mu\nu}}$ 
where $F_{\mu\nu}=\partial_\mu A_\nu - \partial_\nu A_\mu$ and $A_\mu$
is a vector field of dimension 1. But $A_\mu$ is not itself conformal
(the conformally invariant 2-point function of a vector field of
dimension 1 is a gradient, so the associated field strength tensor
would vanish). Although $A_\mu$ has an indefinite 2-point function
($-\eta_{\mu\nu}\cdot (2\pi)^{-2}(x^2)\inv$ in the Feynman gauge),
$F_{\mu\nu}$ is well-known to be a conformally covariant and Wightman
positive free tensor field.

The computations of the 4-point functions are straight forward. We
just give the result in the form of a table, displaying the value for
the amplitude $B_\LL$ of the 2-point function and the values
of the relative amplitudes $\bfa = (a_0,a_1,a_2,b,b')$ according to
(\ref{eq2.1}):  
Our convention for the normalization of fields in the $z$-picture is
such that it includes, apart from the conformal transformation factor,
an additional explicit factor of $2\pi$ so that, e.g., the scalar
field with canonical 2-point function $(2\pi)^{-2}(x_{12}^2)\inv$ in
$x$-space has the 2-point function $(z_{12}^2)\inv$ in $z$-space.

\medskip

\begin{tabular}{lcccccc}
$\LL$ & $B_\LL$ & $a_0$ & $a_1$ & $a_2$ & $b$ & $b'$ 
\\ \hline\\[-3mm]
$\wick{\varphi^4}$ & $24$ & $16$ & - & - & $36$ & $216$ \\
$\wick{\varphi_1^3}\varphi_2$ & $6$ & $10$ & - & - & $18$ & $54$ \\
$\wick{\varphi_1^2}\wick{\varphi_2^2}$ & $4$ & $8$ & - & - & $18$ & $76$ \\
$\wick{\varphi_1^2}\varphi_2\varphi_3$ & $2$ & $6$ & - & - & $10$ & $42$ \\
$\varphi_1\varphi_2\varphi_3\varphi_4$ & $1$ & $4$ & - & - & $6$ & $24$ \\
\hline\\[-3mm]
$\varphi\;\wick{\bar\psi\psi}$ & $16$  & $1$ & $\frac14$ & - & $\frac14$ & - \\
$\varphi\,(\bar\psi_1\psi_2+h.c.)$ & $32$  & $1$ & $\frac18$ & - & $\frac18$ & - \\
\hline\\[-3mm]
$\frac14\,\wick{F_{\mu\nu}F^{\mu\nu}}$ & $48$ & - & - & $\frac 29$ & - & - \\
$\frac14\,F_{1\mu\nu}F_2^{\mu\nu}$ & $24$ & - & - & $\frac 19$ & - & - 
\end{tabular}

\medskip

Clearly, all entries in this table satisfy the bounds given in
Section~\ref{sec3}.

\subsection{Generalized free fields}

Further free field constructions are possible involving generalized
free fields with integer (scalars) or half-integer (spinors) scaling
dimensions above the unitarity bound. Their Wightman functions are
computed from 2-point functions of the generalized free fields by
the same procedure (i.e., Wick's Theorem) as for Wick products of free
fields. 

Again, we need not consider vector fields, because their dimension
above the unitarity bound of 3 is too big to contribute to a scalar of
dimension 4.

Generalized fields do not arise from a Lagrangian, and consequently
there is no associated canonical stress-energy tensor as a Wightman
field. Note, however, that generalized free fields do possess a
stress-energy tensor which is more singular than a Wightman field
\cite{DR03}: it has infinite vacuum fluctuations, hence formally
$c=\infty$; in particular, this stress-energy tensor does not 
contribute to the operator product expansion. 

Moreover, generalized free fields do not possess a canonical
normalization. We are thus free to fix their normalizations by
choosing the 2-point functions in $x$-space to be
$(2\pi)^{-2}(x^2)^{-d}$ for scalars and
$(2\pi)^{-2}i{\not\!\partial}(x^2)^{\frac12-d}$ for spinors.  
As before, we absorb the factor $2\pi$ in the passage to the $z$-space
fields. 

Denoting by $A\equiv \varphi$, $B$, $C$, and $D$ the scalar fields of
dimension 1, 2, 3, 4, and by $\psi$ and $\chi$ spinors of
dimension $\frac32$ and $\frac52$, respectively, we have computed the
following table. 

\medskip
\begin{tabular}{lcccccc}
$\LL$ & $B_\LL$ & $a_0$ & $a_1$ & $a_2$ & $b$ & $b'$ 
\\ \hline\\[-3mm]
$D$ & $1$ & - & - & - & - & - \\ 
$AC$ & $1$ & $1$ & - & - & - & - \\ 
$\wick{B^2}$ & $2$ & - & - & - & $4$ & $8$ \\ 
$B_1B_2$ & $1$ & - & - & - & $2$ & $4$ \\ 
$\wick{A^2}B$ & $2$ & $4$ & - & - & $2$ & $8$ \\ 
$A_1A_2B$ & $1$ & $2$ & - & - & $2$ & $6$ \\ 
\hline\\[-3mm]
$\bar\chi\psi + h.c.$ & $64$ & - & $\frac 1{16}$ & - & - & - \\
\end{tabular}
\medskip

Also the entries of this table satisfy the bounds given in Sect.\ 3,
because generalized free fields with dimension above the unitarity
bounds satisfy Wightman positivity. On the other hand, there is no 
stress-energy tensor (with finite $c$) in all these models. 

\subsection{Subcanonical free fields}
We have also considered free field constructions with $\LL$ a Wick product
involving subcanonical generalized free fields, i.e., generalized free
tensor fields whose dimensions are below the unitarity bound such that these
fields for themselves violate Wightman positivity. But we cannot
exclude a priori the possibility that their Wick products might define
a positive definite subtheory.

With these constructions, there are three possible scenarios how
positivity might fail. 

First, the operator product expansion of $\LL$ with itself may contain
fields whose scaling dimensions are below the unitarity
bound. This occurs with Wick products involving subcanonical Fermi
fields of dimension $\frac 12$, or subcanonical vector fields of
dimension 1, such as $\LL=j^\mu A_\mu$ where
$j^\mu=\wick{\bar\psi\gamma^\mu\psi}$ is a Dirac current and $A_\mu$ a 
subcanonical free conformal vector field of dimension 1. The
operator product expansion of $\LL\LL$ contains the rank 2 symmetric
tensor $\wick{A_\mu A_\nu}-$ trace, whose dimension 2 is below the
unitarity bound. In these cases, there contribute to the 4-point
function partial waves which are more singular than the polynomial
structures $J_\nu$, $D$, $Q$ of Section~\ref{sec2}, and which were
already excluded by the previous analysis \cite{NST03}. We have not
pursued these cases any further.  

Second, the operator product expansion of $\LL$ with itself may contain
only fields with allowed quantum numbers, but a failure of Wightman
positivity at the 4-point level could become manifest through some
negative partial wave coefficient. Remarkably, we did not find
examples for this scenario. 

The third and most intriguing scenario is encountered \cite[work in
progress]{S-un} with the scalar Wick square $\LL=\wick{B_\mu B^\mu}$ of a
subcanonical vector field of dimension 2. Here the 4-point function
is given by the relative amplitudes 
$\bfa=(a_0,a_1,a_2,b,b')=(0,0,\frac14;\frac14,\frac12)$. Remarkably,
all partial wave coefficients are positive, thus Wightman positivity
is satisfied at the 4-point level. Moreover, the field $S_{\mu\nu}$
given by the traceless Wick square of the vector field, which
contributes to the operator product expansion of $\LL$ with itself,
has all the properties of a stress-energy tensor.  
But a careful analysis by separating singular contributions in successive
operator product expansions of $\LL$ shows that positivity-violating
fields such as the bilocal field $\wick{B_\mu(x)B_\nu(y)}$ arise within
the OPE of six $\LL$'s, so Wightman positivity must fail not later
than at the 12-point level. In particular, this model illustrates that
positivity at the 4-point level is certainly not a sufficient
condition for Wightman positivity to all orders.

To conclude, we did not discover any free field model involving
subcanonical conformal fields, which would satisfy positivity to all orders.
The scenario of the free Maxwell tensor being positive to all orders
although it involves a positivity-violating gauge field, does not
repeat itself with subcanonical conformal fields.

\section{Partial wave expansion of scalar four-point functions}\label{appB}
\setcounter{equation}{0}
We describe a method to obtain the partial wave expansion for any
globally conformal invariant scalar 4-point function. We concentrate,
however, on 4-point functions of the form
\bea \label{eqB.1}
\bra A(z_1)B(z_2)B(z_3)A(z_4)\ket = 
 \frac{f(s,t)}{(z_{14}^2)^{d_A} (z_{23}^2)^{d_B}} \equiv 
\frac{(s/t)^{d_A}f(s,t)}
{(z_{12}^2)^{d_A}(z_{23}^2)^{d_B-d_A}(z_{34}^2)^{d_A}} 
\eea
which are relevant for the issue of Wightman positivity. We have used 
this method in order to compute the partial wave coefficients 
(\ref{eq2.8}--\ref{eq2.10}) of the 4-point function (\ref{eq1.5}) of the field 
$\LL$, and to obtain the positivity conditions (\ref{eq5.24}) for the mixed 
4-point function (\ref{eq5.23}). 

Let $d=d_A$ and $\delta=d_B-d_A$. Then the partial wave  
expansion has the form  
\bea
\label{eqB.2}
\bra A(z_1)B(z_2)B(z_3)A(z_4)\ket 
= \frac {1}{(z_{12}^2)^d(z_{23}^2)^\delta(z_{34}^2)^d} \sum B_{\kappa L}\cdot 
\beta_{\kappa L}^\delta(s,t)\,,
\eea
i.e.,
\bea 
\label{eqB.3}
(s/t)^d \cdot f(s,t) = \sum B_{\kappa L}\cdot 
\beta_{\kappa L}^\delta(s,t)\,.
\eea
The partial waves $\beta_{\kappa L}^\delta$ are represented \cite{DO01} as 
\bea 
\label{eqB.4}
\beta_{\kappa L}^\delta(s,t) = 
\frac{uv}{u-v}\; \Big(G_{\kappa+L-\delta/2}^\delta(u)\cdot
G_{\kappa-1-\delta/2}^\delta(v) - (u\leftrightarrow v)\Big) 
\eea
or equivalently
\bea \label{eqB.5}
(s/t)^{\delta}\;\beta_{\kappa L}^\delta(s,t) = 
\frac{uv}{u-v}\; \Big(G_{\kappa+L+\delta/2}^{-\delta}(u)\cdot
G_{\kappa-1+\delta/2}^{-\delta}(v) - (u\leftrightarrow v)\Big) 
\eea
where $u$ and $v$ are the ``chiral variables'' 
\bea \label{eqB.6}
s=uv\,,\qquad t=(1-u)(1-v)\,,\eea
and
\bea \label{eqB.7}
G_\nu^\delta(z) = z^\nu\cdot F(\nu,\nu;2\nu+\delta;z). 
\eea
(For odd $\delta$, only odd twists arise, corresponding to half-integer 
values of $\kappa$, thus $\nu$ is always an integer.) These representations 
of the scalar twist $2\kappa=0,1,2$ partial waves (which are present only 
if $\delta=\pm2\kappa$) involve singular hypergeometric functions 
$F(\cdot,\cdot;2\kappa-2;\cdot)$; but the singularity disappears upon 
antisymmetrization $u\leftrightarrow v$. In fact, these partial waves are 
particularly simple: 
\bea \label{eqB.8}
\beta^{+2\kappa}_{\kappa 0}=1\,, 
\qquad \beta^{-2\kappa}_{\kappa 0}=(s/t)^{2\kappa}\,.  
\eea

\noindent{\em Remark B.1:} Different authors \cite{DO01,LR93} use
different normalizations of the partial waves. Our normalization of
$\beta_{\kappa L}^\delta$ is such that the leading term in $s$ equals
$s^{\kappa-\delta/2}(1-t)^L\cdot (1+O(1-t))$.  

\medskip

Inserting (\ref{eqB.4}), the partial wave expansion of a given 4-point
function (\ref{eqB.1}) is equivalent to the expansion 
\bea \label{eqB.9}
\frac{u-v}{uv} (s/t)^{d}\cdot f(s,t) = \sum B_{\kappa L} 
\Big(G_{\kappa+L-\delta/2}^\delta(u)\cdot
G_{\kappa-1-\delta/2}^\delta(v) - (u\leftrightarrow v)\Big).
\eea

For GCI 4-point functions, the function $f(s,t)$ is rational, in fact,
a finite linear combination of terms of the form $s^rt^q$, where
$r > -d$ except for a possible distinguished term $(s/t)^{-d}$ which
can be present only if $\delta \geq 0$. The latter corresponds separately to
the scalar partial wave $\beta^{\delta}_{\kappa=\delta/2,L=0}$ (= the
vacuum contribution if $A=B$). It is then possible to represent
the left-hand side of (\ref{eqB.9}) (without the distinguished term)
as a finite linear combination of terms of the form  
\bea \label{eqB.10}
\hbox{(nonnegative power of $u$ or $\frac u{1-u}$)} \cdot 
\hbox{(nonnegative power of $v$ or $\frac v{1-v}$)}  - (u\leftrightarrow v)\,.
\eea

\noindent {\em Remark B.2:} The 4-point structures $J_\nu$, $D$ and
$Q$ of Sect.\ \ref{sec2} simplify considerably when represented in
this way in terms of the chiral variables $u$ and $v$.  
E.g., the twist 2 contributions $j_\nu$ of eq.\ (\ref{eq2.4}), after
multiplication by $(u-v)$, become simply $[(-1)^\nu u^{\nu+1} + (\frac
u{1-u})^{\nu+1}] - (u\leftrightarrow v)$. 
\medskip

Each term in (\ref{eqB.10}) has to be expanded into products 
$(G_\mu^\delta(u)G_\nu^\delta(v) -
(u\leftrightarrow v))$. This is achieved with the help of the
expansion formulae 
\bea \label{eqB.11}
z^p = \sum_{\nu\in p+\NN_0}\frac{(-1)^{\nu-p}}{(\nu-p)!}
\frac{(p+\alpha)_{\nu-p}(p+\beta)_{\nu-p}}{(\nu+p+\gamma-1)_{\nu-p}}
\cdot z^\nu F(\nu+\alpha,\nu+\beta;2\nu+\gamma;z), 
\eea
\vskip-2mm
\bea \label{eqB.12}
\left(\frac z{1-z}\right)^p = 
(1-z)^\alpha\sum_{\nu\in p+\NN_0}\frac{1}{(\nu-p)!}
\frac{(p+\alpha)_{\nu-p}(p+\gamma-\beta)_{\nu-p}}{(\nu+p+\gamma-1)_{\nu-p}}
\cdot z^\nu F(\nu+\alpha,\nu+\beta;2\nu+\gamma;z)
\eea 
which are valid if $2p+\gamma>0$. We shall prove these expansions below.

If $\delta\geq0$, we may choose\footnote{Other instances of these
  formulae are relevant for the partial wave expansion of mixed
  4-point functions involving scalar fields of different dimensions.}  
$\alpha=\beta=0$, $\gamma=\delta$ and apply (\ref{eqB.11}),
(\ref{eqB.12}) to both factors in (\ref{eqB.10}). 
Thus, the left-hand side of (\ref{eqB.9}) (without the distinguished
term) is represented as a series of the form $\sum_{\mu\nu} X_{\mu\nu}
(G_\mu^\delta(u)G_\nu^\delta(v) - (u\leftrightarrow v))$, with $\mu
\neq \nu$ running over all nonnegative integers. It remains to
relabel $\mu=\kappa+L-\delta/2$ and $\nu=\kappa-1-\delta/2$ if
$\mu>\nu$, and $\mu=\kappa-1-\delta/2$ and $\nu=\kappa+L-\delta/2$ if
$\mu<\nu$. Comparing with (\ref{eqB.9}), one obtains the partial wave
coefficients  
\bea \label{eqB.13}
B_{\kappa L} = X_{\kappa+L-\delta/2,\kappa-1-\delta/2} -
X_{\kappa-1-\delta/2,\kappa+L-\delta/2}\qquad (2\kappa>\delta)\,.
\eea

If $\delta<0$, the expansions (\ref{eqB.11}) and (\ref{eqB.12}) for
small $p$ involve singular hypergeometric functions with
$2\nu+\delta<0$. To avoid this problem, one proceeds in the same way
by using the alternative representation (\ref{eqB.12}) instead of
(\ref{eqB.4}). Equivalently, one may use the the ``reversed'' 4-point
function $\bra BAAB \ket$, which has the same partial wave
coefficients as $\bra ABBA \ket$. 

\noindent{\em Remark B.3:} If $A=B$, then locality implies the
symmetry of (\ref{eqB.1}) under the permutation $(z_1\leftrightarrow z_2)$. 
It follows that (\ref{eqB.9}) is antisymmetric under $u\mapsto
-u/(1-u)$ and $v\mapsto -v/(1-v)$, and consequently the symmetry
$G_\nu^0(z) = (-1)^{\nu}\, G_\nu^0(-\frac z{1-z})$ cancels all terms
with odd $L$, thus reproducing the well-known fact that the operator
product expansion of a field with itself produces only fields of even spin. 

Suitably adapting the parameters $\alpha,\beta,\gamma$, the
method works also for the general mixed 4-point function. 

\medskip

It remains to prove (\ref{eqB.11}) and (\ref{eqB.12}). The latter is
equivalent to the former by the functional identity for hypergeometric
functions under $z\mapsto -z/(1-z)$. By shifting both the summation
index and the parameters $\alpha,\beta,\gamma$ by $p$, (\ref{eqB.11}) is
equivalent to the identity ($\gamma>0$) 
\bea \label{eqB.14} 
1 = \sum_{n=0}^\infty
\frac{(-1)^n}{n!}\frac{(\alpha)_n(\beta)_n}{(\gamma+n-1)_n}
\cdot z^n F(n+\alpha,n+\beta;2n+\gamma;z).\eea
To prove this identity, we insert the expansion for the hypergeometric
functions and collect the coefficients of $z^k$, giving
$$
\frac{(\alpha)_k(\beta)_k}{k!}\sum_{n=0}^k(-1)^n\cdot (\gamma+2n-1)
\left(k\atop n\right)\cdot \frac{1}{(\gamma+n-1)_{k+1}}.
$$
This sum trivially equals 1 if $k=0$. For $k>0$, we insert the
elementary identity $(\gamma+2n-1) = (\gamma+k+n-1)\cdot $ $\frac nk +
(\gamma+n-1)\cdot \frac{k-n}k$, giving  
$$
\frac{(\alpha)_k(\beta)_k}{k!}\sum_{n=0}^k(-1)^n\cdot \left[
\left(k-1 \atop
  n-1\right) \frac{1}{(\gamma+n-1)_{k}} + \left(k-1\atop
  n\right)\frac{1}{(\gamma+n)_{k}}\right] = 0.
$$
This proves (\ref{eqB.14}), hence (\ref{eqB.11}) and (\ref{eqB.12}). 

\section{Six-point twist two contributions}\label{appC}
\setcounter{equation}{0}

We shall sum up results about the 6-point function
\bea
\label{eqC1}
F(1,2;3,4;5,6) := \bra V_1 (z_1,z_2) \, V_1 (z_3,z_4) \, V_1 (z_5,z_6) \ket 
\eea
needed in Section~\ref{sec5}. The function $F$ appears as the leading
contribution for $z_{2k-1,2k}^2 \to 0$, $k = 1,2,3$ to the truncated
6-point function $w_6^t$ of $\LL(z)$ 
\bea
\label{eqC2}
w_6^t (z_1 , \ldots , z_6) = B^3 (z_{12}^2 \, z_{34}^2 \,
z_{56}^2)^{-3} \{ F (1,2;3,4;5,6) + O (\vert z_{12}^2 \vert + \vert
z_{34}^2 \vert + \vert z_{56}^2 \vert)\} 
\eea
and is characterized by the following properties: (i) GCI; (ii) $F$ is
a rational function of $z_{ij}^2$ ($i < j$) with not higher than third
degree poles and no singularities in $z_{2k-1,2k}^2$, $k = 1,2,3$;
(iii) it is invariant under the 48 element subgroup $G_{8,6}$ of the
permutation group ${\mathcal S}_6$ of $\{ z_1 , \ldots , z_6 \}$
generated by the substitutions $s_{2k-1,2k}$, $k = 1,2,3$ and the
products $s_{13} \, s_{24} , \, s_{35} \, s_{46}$ which exchange pairs
of variables ($(z_1 , z_2) \rightleftarrows (z_3 , z_4)$ etc.); (iv)
$F$ is harmonic in each $z_i$. We shall denote the linear manifold of
all functions satisfying (i)--(iv) by ${\mathcal H}_6$. 

\smallskip

It turns out that the space ${\mathcal H}_6$ is eight-dimensional. A basis
in it, -- {\it i.e.} a set of eight linearly independent rational
functions satisfying the above conditions, will be displayed elsewhere. 
Here we shall briefly describe the method of deriving the general
expression for $F$ and shall write down the subset of four linearly
independent functions for which $V_1 (z,z) \ne 0$, implying the
existence of a (hermitian) scalar field $\phi$ of dimension 2. Of
the remaining four we shall display the one with lowest order of
singularity (2) in $z_{ij}^2$.  

\smallskip

We start with an observation which singles out GCI $n$-point functions
with $n \leq 6$ in four space-time dimensions. 

\smallskip

Consider the commutative algebra ${\mathcal A}^{(n)}$ of $n(n-1)$
generators, $\rho_{ij}$ and $\rho_{ij}^{-1}$ for $1 \leq i < j \leq
n$. Let ${\mathcal A}^{(n,{\mathcal D})}$ be the algebra of scalar
$n$-point functions in $D$-dimensional (compactified) space-time
$\overline \MM$ spanned by monomials of the type $\underset{1 \leq i < j \leq
  n}{\prod} (z_{ij}^2)^{\mu_{ij}}$ with $\mu_{ij} \in \ZZ$,
$z_{ij} = z_i - z_j$. For $n \geq 3$ the homomorphism $J_D : {\mathcal
  A}^{(n)} \to {\mathcal A}^{(n,{\mathcal D})}$ such that $\rho_{ij}
\to z_{ij}^2$ is an isomorphism for $(3 \leq) \, n \leq
D+1$. Otherwise, for $n > {\mathcal D} + 1$, there are
$\begin{pmatrix} n-D \\ 2 \end{pmatrix}$ relations coming from the
vanishing of Gram determinants of inner products of linearly dependent
vectors. Thus, for space-time dimension $D=4$ and $n \leq 5$ the map 
$J_D$ is an
isomorphism. We shall now demonstrate that this allows to construct a
complete set of {\it GCI 6-point functions} treating $\rho_{ij} =
z_{ij}^2$ as independent, in spite of the fact that there is a
(single, degree five) relation among them in this case. 

\smallskip

We first note that ${\mathcal A}^{(n)}$ (as well as ${\mathcal
  A}^{(n,{\mathcal D})}$) admits a $\ZZ^n$ grading ${\mathcal
  A}^{(n)} = \underset{d_1 , \ldots , d_n}{\bigoplus} {\mathcal
  A}^{(n)}_{d_1 \ldots d_n}$ defined by 
\bea
\label{eqC3}
\hbox{weight} \ \left( \prod_{1 \leq i < j \leq n}
  \rho_{ij}^{\mu_{ij}} \right) = (d_1 , \ldots , d_n) \in \ZZ^n \quad
\hbox{for} \quad d_k = - \sum_{i=1}^{k-1} \mu_{ik} - \sum_{j=k+1}^n
\mu_{kj} \, , 
\eea
the allowed weights satisfying
\bea
\label{eqC4}
\sum_{k=1}^n d_k \left( =-2 \sum_{i < j} \mu_{ij} \right) \in 2 \,
\ZZ \, . 
\eea
It is a crucial observation that the GCI $n$-point function $\bra
\phi_1 (z_1) \ldots \phi_n (z_n) \ket$ where $\phi_i (z)$ is a
conformal scalar field of dimension $d_i$ is an $n$-homogeneous
element of ${\mathcal A}^{(n,{\mathcal D})}$ of weight $(d_1 , \ldots
, d_n)$. This follows from the fact that under a conformal
transformation $z \to g(z)$ the monomial in the left hand side of
(\ref{eqC3}) transforms as 
\bea
\label{eqC4bis}
\prod_{1 \leq i < j \leq n} [(g(z_i) - g(z_j))^2]^{\mu_{ij}} =
\prod_{k=1}^n \ [\omega (g , z_k)]^{d_k} \prod_{1 \leq i < j \leq n}
(z_{ij}^2)^{\mu_{ij}} \, , 
\eea
where $\omega (g,z)$ is a quadratic polynomial in $z$. As one can send
any point to infinity by an appropriate (complex) conformal
transformation, every $(n+1)$-homogeneous monomial of weight $(d_0 ,
d_1 , \ldots , d_n)$ 
\bea
\label{eqC5}
\prod_{0 \leq i < j \leq n} \rho_{ij}^{\mu_{ij}} \in {\mathcal A}_{d_0
  d_1 \ldots d_n}^{(n+1)} 
\eea
is uniquely determined by the factor
\bea
\label{eqC6}
\prod_{1 \leq i < j \leq n} \rho_{ij}^{\mu_{ij}} \in {\mathcal
  A}_{\delta_1 \ldots \delta_n}^{(n)} \quad \hbox{where} \quad
\delta_1 + \ldots + \delta_n = d_1 + \ldots + d_n - d_0 \, , 
\eea
independent of the variable $z_0$. Indeed, given $(\delta_1 , \ldots ,
\delta_n)$ satisfying (\ref{eqC6}), one can restore (\ref{eqC5}) by
setting $\mu_{0j} = \delta_j - d_j$. In particular, the set of GCI
6-point functions is determined by the set of homogeneous (rational)
5-point functions for which the variables $\rho_{ij}$ are
independent. 

\smallskip

Our next task is to find a basis of 6-homogeneous harmonic functions
of weight $(1,1,1,1,1,1)$ constraint by condition (ii) on their
singularities. This is made easy by the fact that the Laplace operator
acting on ${\mathcal A}_{111111}^{(6,4)}$ acquires a simple form in
terms of the variables $\rho_{ij}$: if we set $\rho = \{ \rho_{k\ell}
, 1 \leq k < \ell \leq 6 \}$, $F (1,2 , 3,4 ; 5,6) = f(\rho)$ (or,
using the above homomorphism $J_{\mathcal D}$ for ${\mathcal D} = 4$,
$F = J_4 \, f$) we find 
\bea
\label{eqC7}
\partial_{z_1}^2 \, F (1,2;3,4;5,6) = - {\mathcal D}_1 \, f (\rho) = 0 \, ,
\quad {\mathcal D}_1 := \sum_{1 < i < j \leq 6} \rho_{ij} \,
\frac{\partial^2}{\partial \rho_{1i} \, \partial \rho_{1j}} \, . 
\eea
Here we have used the Euler equation $\overset{6}{\underset{k=2}\sum}
\rho_{1k} \, \frac{\partial}{\partial \rho_{1k}} \, f = -f$. 

\medskip

\noindent {\bf Lemma C.1: \it If $f = \underset{\mu}{\sum} \ C(\mu)
  \underset{1 \leq i < j \leq 6}{\prod} \rho_{ij}^{\mu_{ij}}$, $(\mu
  \in \ZZ^{15}$ being a multiindex), is a solution of
  (\ref{eqC7}) then 
\bea
\label{eqC8}
C(\mu) = 0 \ \hbox{if there exist} \ 1 < k < \ell \ \hbox{such that} \ \mu_{1k} < 0 \ \hbox{and} \ \mu_{1\ell} < 0 \, .
\eea
}

To {\it prove} the lemma one chooses the minimal ({\it i.e.}\ the
largest in absolute value, negative) $\mu_{1k}$ and $\mu_{1\ell}$
appearing in a single term in the expansion of $f$ and then concludes
that the derivative $\rho_{k\ell} \, \frac{\partial^2}{\partial
  \rho_{1k} \, \partial \rho_{1\ell}}$ of this monomial cannot be
cancelled by any other term in the sum. 

\smallskip

After these preliminaries, writing down a basis in ${\mathcal H}_6$,
{\it i.e.}\ a complete set of linearly independent rational functions
satisfying conditions (i)--(iv), becomes a (more or less) routine
technical problem. 

\smallskip

We shall present the solution in increasing order of singularity
starting with first order poles. According to Lemma~B.1 and to the
symmetry assumption (iii) there are eight possible singular
configurations (for a given order of singularity $n$), obtained from
$(\rho_{16} \, \rho_{23} \, \rho_{45})^{-n}$ by transposing
independently $1  \rightleftarrows 2$, $3  \rightleftarrows 4$, $5
\rightleftarrows 6$. Each such {\it elementary contribution} (with a
fixed pole structure) is symmetric under the trihedral subgroup
${\mathcal D}_3$ ($\sim {\mathcal S}_3$) of $G_{8,6}$ generated by the
(order three) cyclic permutation $(1 \, 2 \, 3 \, 4 \, 5 \, 6) \to (5
\, 6 \, 1 \, 2 \, 3 \, 4)$ and the reflection $(1 \, 2 \, 3 \, 4 \, 5
\, 6) \to (6 \, 5 \, 4 \, 3 \, 2 \, 1)$. Thus, the unique vector in
${\mathcal H}_6$ with only first order pole singularities is 
\bea
\label{eqC9}
F_1 (1,2;3,4;5,6) = S (\ZZ_2^3) (\rho_{16} \, \rho_{23} \, \rho_{45})^{-1}
\eea
where $S (\ZZ_2^3)$ is symmetrization with respect to the 8-element
abelian group $\ZZ_2^3$ generated by the transpositions
$s_{12} , s_{34}$ and $s_{56}$: 
\bea
\label{eqC10}
S(\ZZ_2^3) (\rho_{16} \, \rho_{23} \, \rho_{45})^{-1} &= &(\rho_{16}
\, \rho_{23} \, \rho_{45})^{-1} + (\rho_{13} \, \rho_{26} \,
\rho_{45})^{-1} + (\rho_{16} \, \rho_{24} \, \rho_{35})^{-1} +
(\rho_{15} \, \rho_{23} \, \rho_{46})^{-1} \nonumber \\ 
&+ &(\rho_{15} \, \rho_{24} \, \rho_{36})^{-1} + (\rho_{13} \,
\rho_{25} \, \rho_{46})^{-1} + (\rho_{14} \, \rho_{26} \,
\rho_{35})^{-1} + (\rho_{14} \, \rho_{25} \, \rho_{36})^{-1} \, . 
\eea
It is reproduced by the normal product of a free massless scalar field
$\varphi$ with itself: $V_1^{\varphi} (z_1 , z_2) = \wick{ \varphi (z_1)
  \, \varphi (z_2)}$.  

There are two linearly independent functions in ${\mathcal H}_6$ with
poles of order not exceeding two: 
\bea
\label{eqC11}
F_2^{(i)} (1,2;3,4;5,6) = S (\ZZ_2^3) \, W_2^{(i)} (12;34;56) \qquad i=1,2
\eea
where
\bea
\label{eqC12}
W_2^{(1)} (12;34;56) &= &\rho_{16}^{-2} \, \rho_{23}^{-2} \,
\rho_{45}^{-2} \, S (\ZZ_3) \, \{ \rho_{16} (\rho_{25} \,
\rho_{34} - \rho_{24} \, \rho_{35}) \nonumber \\ 
&= &\frac{\rho_{16} (\rho_{25} \, \rho_{34} - \rho_{24} \, \rho_{35})
  + \rho_{23} (\rho_{14} \, \rho_{56} - \rho_{16} \, \rho_{45}) +
  \rho_{45} (\rho_{12} \, \rho_{36} - \rho_{13} \,
  \rho_{26})}{\rho_{16}^2 \, \rho_{23}^2 \, \rho_{45}^2} \, , 
\eea
\bea
\label{eqC13}
W_2^{(2)} (12;34;56) &= &\frac{1}{\rho_{16} \, \rho_{23} \, \rho_{45}}
+ \frac{\rho_{12} \, \rho_{34} \, \rho_{56} - \rho_{14} \, \rho_{25}
  \, \rho_{36}}{\rho_{16}^2 \, \rho_{23}^2 \, \rho_{45}^2} \nonumber
\\ 
&+ &S(\ZZ_3) \, \frac{\rho_{23} (\rho_{14} \, \rho_{56} - \rho_{15} \,
  \rho_{46}) - \rho_{13} \, \rho_{24} \, \rho_{56} + \rho_{14} \,
  \rho_{26} \, \rho_{35}}{\rho_{16}^2 \, \rho_{23}^2 \, \rho_{45}^2}
\, , 
\eea
where $S(\ZZ_3)$ stands for symmetrization with respect to the
3-element cyclic group generated by the permutation $(1 \, 2 \, 3 \, 4
\, 5 \, 6) \to (5 \, 6 \, 1 \, 2 \, 3 \, 4)$. The first of them is
characterized by the fact that it has non-zero limit for $z_{2k-1} =
z_{2k}$, $k = 1,2,3$; for instance, 
\bea
\label{eqC14}
W_2^{(1)} (12;33;44) = \frac{\rho_{12} \, \rho_{34} - \rho_{13} \,
  \rho_{24} - 2 \, \rho_{14} \, \rho_{23}}{\rho_{14}^2 \, \rho_{23}^2
  \, \rho_{34}} \, . 
\eea
The function $F_2^{(2)}$ is reproduced by the bilocal field
$V_1^{\psi}$ (\ref{eq4.21}) (the composite of the free Weyl field)
({\it cf.} \cite{NT04}): 
\bea
\label{eqC17}
\bra V_1^{\psi} (z_1 , z_2) \, V_1^{\psi} (z_3 , z_4) \, V_1^{\psi}
(z_5 , z_6) \ket = \frac{1}{8} \, F_2^{(2)} (1,2;3,4;5,6) \, . 
\eea

There are two more independent elements of ${\mathcal H}_6$, both with
third degree poles, also having non-zero limit for $z_{2k-1} =
z_{2k}$, $k = 1,2,3$. Setting again $F_3^{(i)} = S(\ZZ_2^3) \,
W_3^{(i)}$, $i = 1,2$, we find 
\bea
\label{eqC15}
W_3^{(1)} (12;34;56) = S(\ZZ_3) \left\{ \frac{(\rho_{14} \,
    \rho_{56} - \rho_{15} \, \rho_{46})^2}{\rho_{23} \, \rho_{45}^3 \,
    \rho_{16}^3} - \frac{\rho_{15} \, \rho_{46}}{\rho_{23} \,
    \rho_{45}^2 \, \rho_{16}^2} \right\} \, , 
\eea
\bea
\label{eqC16}
W_3^{(2)} (12;34;56) &= &3 \,  \frac{\rho_{12} \, \rho_{34} \,
  \rho_{56}}{\rho_{23}^2 \, \rho_{45}^2 \, \rho_{16}^2} \nonumber \\ 
&+ &S(\ZZ_3) \left\{ \frac{\rho_{14} \, \rho_{26} \, \rho_{35}
    - 2 \, \rho_{13} \, \rho_{24} \, \rho_{56}}{\rho_{23}^2 \,
    \rho_{45}^2 \, \rho_{16}^2} - 2 \,  \frac{\rho_{12} \, \rho_{14}
    \, \rho_{36} \, \rho_{56} + \rho_{13} \, \rho_{15} \, \rho_{26} \,
    \rho_{46}}{\rho_{23}^2 \, \rho_{45}^2 \, \rho_{16}^3} \right\}
\nonumber \\ 
&+ &2 \, S (D_3) \, \frac{\rho_{13} \, \rho_{14} \, \rho_{26} \,
  \rho_{56}}{\rho_{23}^2 \, \rho_{45}^2 \, \rho_{16}^2} \, , 
\eea
where $S(D_3)$ stands for symmetrization with respect to the trihedral
group ${\mathcal D}_3$ described before Eq.~(\ref{eqC9}). None of
these two structures can be reproduced by free fields. 

\smallskip

The remaining three structures (which will be studied elsewhere) all
have third order poles and vanish for coinciding arguments. One of
them is reproduced by the 6-point function of $V_1^F$ (\ref{eq4.29}). 

\smallskip

We note that the 4-point functions $\rho_{13} \, \rho_{24} \,
\rho_{34} \, F_n^{(i)} (1,2;3,3;4,4)$ for $(n,i) = (2,1), (3,1),
(3,2)$ are, as expected, linear combinations of $j_{\nu} (s,t)$. For
instance, $\frac{1}{4} \, \rho_{13} \, \rho_{24} \, \rho_{34} \,
F_2^{(1)} (1,2;3,3;4,4) = - \, 3 \, j_0 (s,t) - j_1 (s,t)$. 


\end{document}